\documentclass{aastex}

\usepackage{emulateapj5}

\begin{document}

\title{Exploring Halo Substructure with Giant Stars IV:  The Extended
Structure of the Ursa Minor
Dwarf Spheroidal}

\author{Christopher Palma\altaffilmark{1}, Steven R. Majewski\altaffilmark{2}, 
Michael H. Siegel\altaffilmark{3,4}, 
Richard J. Patterson\altaffilmark{4},
James C. Ostheimer, \& Robert Link\altaffilmark{5}}

\affil{Department of Astronomy, University of Virginia  \\
       Email:  cp4v@virginia.edu, srm4n@virginia.edu, mhs4p@virginia.edu,
       ricky@virginia.edu, jco9w@virginia.edu, rl8z@virginia.edu}
\authoraddr{P.O.\ Box 3818, Charlottesville, VA  22903-0818}

\altaffiltext{1}{Current Address:  Department of Astronomy \& Astrophysics,
Penn State University, 525 Davey Laboratory, University Park, PA 16802.
-- cpalma@astro.psu.edu}

\altaffiltext{2}{David and Lucile Packard Foundation Fellow; Cottrell Scholar
of the Research Corporation; National Science Foundation CAREER Fellow.}

\altaffiltext{3}{Current Address: Space Telescope Science Institute,
3700 San Martin Drive, Baltimore, MD 21218. -- msiegel@stsci.edu}

\altaffiltext{4}{Visiting Astronomer, Kitt Peak National Observatory,
National Optical Astronomy Observatories, which is operated by the
Association of Universities for Research in Astronomy, Inc. (AURA)
under cooperative agreement with the National Science Foundation.}

\altaffiltext{5}{Current Address:  Northrop Grumman Information Technology - TASC,
4801 Stonecroft Boulevard,
Chantilly, VA 20151}

\begin{abstract}

We present a large area photometric survey of the Ursa Minor dwarf
spheroidal galaxy and its environs.  This survey is intended to trace
the distribution of stars outside the nominal tidal radius of this
system.  Observations were made with the Washington $M$, Washington
$T_2$, and $DDO51$ filters, which in combination have been shown
previously to provide reliable stellar luminosity classification for K
type stars.  We identify giant star candidates with the same distance
and metallicity as known Ursa Minor RGB stars extending to
approximately 3\arcdeg\ from the center of the dSph.  Comparison to
catalogues of stars within the tidal radius of Ursa Minor that have
been observed spectroscopically suggests that our photometric
luminosity classification is 100\% accurate.  Over a large fraction of
the survey area, our photometry is deep enough that blue horizontal
branch stars associated with Ursa Minor can also be identified.  The
spatial distribution of both the candidate Ursa Minor giant stars and
the candidate BHB stars are remarkably similar, and, for both samples,
a large fraction of the stars are found outside the nominal tidal
radius of Ursa Minor.  An isodensity contour map of the surface density
of stars within the tidal radius of Ursa Minor reveals several
morphological peculiarities:  (1)  The highest density of dSph stars is
not found at the center of symmetry of the outer isodensity contours,
but instead is offset southwest of center.  (2)  The overall shape of
the outer contours does not appear to be elliptical, but appears
\textsf{S}-shaped.  A surface density profile was derived for Ursa
Minor and compared to those derived from previous studies.  We find
that previously determined King profiles with $\sim50\arcmin$ tidal
radii do not fit well the distribution of candidate UMi stars
identified in this study, which extends to greater radii than these
other surveys.  A King profile with a much larger tidal radius produces
a reasonable fit, however a power law with index $-3$ provides an even
better fit to the densities at radii greater than $20\arcmin$.  The
existence of Ursa Minor associated stars at large distances from the
core of the galaxy, the peculiar morphology of the galaxy within its
tidal radius, and the shape of its surface density profile all suggest
that this system is evolving significantly due to the tidal influence
of the Milky Way.  However, the photometric data on Ursa Minor stars
alone do not allow us to determine if the candidate extratidal stars
are now unbound or if they remain bound to the dSph within an extended
dark matter halo.

\end{abstract}

\keywords{galaxies: evolution --- galaxies: formation --- galaxies: halos ---
galaxies: individual (Ursa Minor dSph) --- galaxies: photometry --- 
galaxies: structure}

\section{Introduction}

The study of outer halo objects as tracers of possible substructure of
the outer halo is motivated primarily by the Galactic halo formation
scenario of \citet[][hereafter, SZ]{SZ}.  In contrast to the Galactic
formation model of \citet{els}, which postulates that the Galaxy formed
during a single, rapid collapse of a proto-Galactic cloud, SZ argue
that the outer halo of the Galaxy may be made up primarily of stars and
stellar systems that formed in transient ``fragments'' that have
subsequently fallen into and been accreted by the Galaxy after the
major collapse phase had been completed.  While the SZ model for the
formation of the outer halo derives solely from observations of outer
halo globular clusters, in the quarter century since the SZ model was
introduced, direct observations of outer halo stars have provided
support for an accretion origin of at least some of the outer halo.
For example, the implied age spread between inner halo and outer halo
field blue horizontal branch stars \citep{psb91}, the measurement of a
net retrograde rotation for certain volumes of halo stars
\citep{srm92a}, and the existence of stellar ``moving groups''
\citep[e.g.,][]{db89, srmetal94, srmetal96} are all plausible
manifestations of the infall and accretion of distinct stellar systems
by the Milky Way.

Moreover, the accumulating evidence that the Sagittarius dwarf galaxy
\citep{i1, ib95} has left stars and globular clusters strewn along its
orbit \citep{mmetal98, srm99, dd00, sdss, palma02, ibcarb} provides
direct support for the \citet{SZ} accretion hypothesis.  N-body
simulations of the evolution of dwarf spheroidal galaxies (dSphs) in
the gravitational field of the Milky Way \citep[e.g.,][]{mcg90,
moore94, oh, kvj96} predict that the tidal disruption of dSphs may
occur, and that these galaxies are likely to contribute stars to the
outer halo in coherent streams\footnote[1]{We note that the
\citet{helmi} investigation of tidal debris in the {\em inner} halo
finds that stars initially on stream-like orbits will not retain any
spatial coherence after a relatively short time, but in order to
conserve phase space density, will become increasingly dynamically
coherent.}.  Recently, significant observational efforts have been
devoted to the search for unbound stars associated with other Galactic
dSphs \citep{ih95, ksm96, haloII, kkaas, ppao01, md01, sdssdra} to determine if
extratidal stars are a ubiquitous feature in these galaxies, or whether the
disruption of Sagittarius is a unique event in the Milky Way's history.

The study of the internal dynamics and dark matter content of the dSphs
is complicated by the unknown extent that the effects of ongoing tidal
disruption may have on the equilibrium state of the systems.  The
central velocity dispersions of the dSph satellites of the Milky Way
are all $\gtrsim7$ km/sec \citep{mm98}.  Assuming that (1) mass follows
light, (2) the internal velocity dispersion is isotropic, and (3) the
system is in virial equilibrium, these large velocity dispersions imply
that the mass-to-light ratio ($M/L$) for the dSphs range up to $\sim
100$, an upper limit defined by Draco and Ursa Minor \citep{har94,
taft95, mm98}.  If these $M/L$ values are correct, the Galactic
satellite dSphs have the highest fractional dark matter content of any
known stellar system.

Because such large $M/L$ values are controversial, it has often been
suggested that the velocity dispersions for the dSphs may be inflated
by some mechanism.  For example, the inclusion of a number of
spectroscopic binaries in the sample of stars used to derive the
velocity dispersion may inflate the value.  However, multiple epoch
observations of binaries in Ursa Minor and Draco \citep{opa96} and
simulations of the effects of binary stars on the velocity dispersions
of dSphs \citep{har96} find that this effect is likely minimal.
\citet{kuhn89} proposed another alternative; the varying tidal force
felt by a dSph on an elliptical orbit may excite resonances in the
galaxy, heating the stars and inflating the velocity dispersion.
However, \citet{pry96} disagrees; he claims that this mechanism can not
inflate the dispersion by the factor of 10-20 claimed by
\citet{kuhn89}, and also that the \citet{kuhn89} resonance would create
a velocity gradient in the dSph that is not seen in observations of
Ursa Minor and Draco.

Although several theoretical
studies cast doubt on the assertion that tidal disruption may be
responsible for the large $M/L$ values for the dSphs \citep{oh, pp95,
kvj99b}, at least one simulation proposes a model where a dSph with no
dark matter can have an observed $M/L$ of nearly 100 \citep{kk98}.  In
the high resolution simulations of \citet{kk98}, a stellar system is
followed for many orbits, until it is nearly completely disrupted,
which is in contrast to previous models that were not only of lower
resolution but were only followed for short interaction timescales.
The remnant of this disruption contains 1\% of the initial satellite
mass and has properties similar to those observed for the present day
dSphs.  Furthermore, if the \citet{kk98} remnants are observed with
favorable orientations, the central velocity dispersion is large even
though the system contains no dark matter.   Observations suggest that
we are unlikely to be seeing systems like Ursa Minor in the orientation
required by this particular family of models, however these models do
demonstrate that it is possible for tidally disrupted stellar systems
without dark matter to exhibit inflated $M/L$ values.

Since the identification of tidal tails associated with dSphs bears on
both the formation and structure of the Galaxy as well as the structure
and evolution of the dark matter halos of dwarf satellite galaxies, we
have undertaken a targeted search of the environments around dSph
satellites of the Milky Way  as one tactic in our exploration of ``halo
substructure'' \citep{haloI, haloII}.  This paper details the extension
of this campaign to the Ursa Minor dwarf spheroidal galaxy.

The study of Ursa Minor is particularly germane to the study of the
Galactic tidal effects on dwarf satellite galaxies;  Ursa Minor has
long been suspected of experiencing ongoing tidal disruption.
\citet{hodge69} concluded that Ursa Minor may be ``broken up due to the
strong galactic tidal force''.  Ursa Minor's morphology, particularly its large
ellipticity and clumpy stellar distribution \citep[e.g.,][]{oa85, ih95,
demers95, schesk01}, are indicative of a system far from relaxation.
Ursa Minor is also crucial to resolving the controversy over the
possible effects tidal disruption may have on internal dSph velocity dispersions
since it lies at the extreme end of the range of predicted dark matter
content by virtue of its large inferred $M/L$ ratio and its faint luminosity
\citep{mm98}.

In this paper, we present a large area photometric survey
of the Ursa Minor (UMi) dSph and its environs (\S2) and discuss in
detail the selection of candidate Ursa Minor member stars (\S3).  With
our sample of Ursa Minor stars, we analyze the spatial distribution and
surface density profile of the galaxy (\S4).  We conclude that Ursa
Minor is surrounded by a significant population of extratidal stars
(\S5), and discuss the implications of this extended population.

There are several previous large area photometric surveys of Ursa Minor
useful as benchmarks for comparison to the survey presented here.  The
two largest surveys are those of \citet[][hereafter, IH95]{ih95} and
\citet[][hereafter, K98]{kleyna98}.  The former survey relied on
$6\arcdeg \times 6\arcdeg$ photographic plates of Ursa Minor taken with
the Palomar Schmidt telescope.  The plates were digitized and an
isopleth map of the dSph was created using the stars found within a
$3\arcdeg \times 3\arcdeg$ degree region centered on the galaxy.  A
background stellar density was calculated and subtracted off of the
surface density profile for the galaxy, which was calculated from the
isopleth map.  The latter survey consists of a grid of 27 overlapping
$V-$ and $I-$band CCD images of Ursa Minor covering somewhat less than
$1\arcdeg \times 1\arcdeg$.  Candidate Ursa Minor stars were separated
from the field population using color information.  This color
selection reduces the background level in the K98 study by
approximately a factor of 4 compared to the IH95 survey.  The
survey presented here complements both of these previous, large surveys
in that it covers approximately the same amount of area as
IH95, and further reduces the background contamination compared
to the deep survey of K98.

\section{Observations and Photometry}

\begin{figure}
\epsscale{0.75}
\plotone{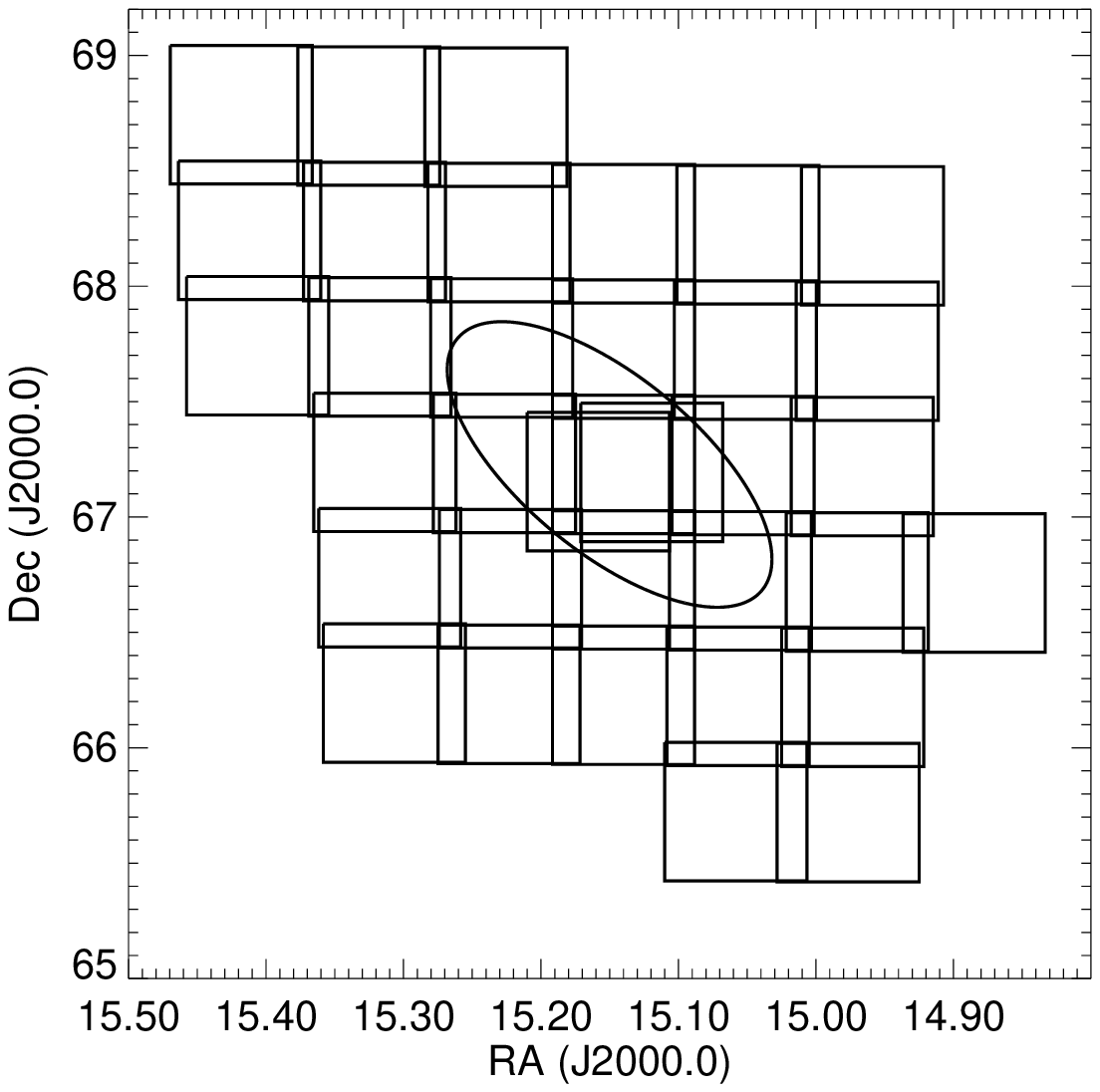}
\caption{The grid of Mosaic frames observed around Ursa Minor.  Three
overlapping frames with longer exposure times were taken of the core
of the galaxy.  Surrounding the core fields, an array of 32 additional
frames were observed with shorter exposure times.  The approximate 
tidal radius of Ursa Minor is shown as an ellipse for reference.}
\label{fig:grid}
\end{figure}

Observations of Ursa Minor were obtained with the Mayall 4 meter
telescope at Kitt Peak National Observatory on the nights of 24 to 26
May 1999.  The detector in use was the Mosaic camera, a $4 \times 2$
array of $2048 \times 4096$ pixel CCDs, which provides a $36\arcmin
\times 36\arcmin$ field of view.  Data on Ursa Minor were taken during
photometric conditions.  A grid of survey fields was arranged such that
adjacent fields overlap by $5\arcmin$:  Three dithered sets of frames
were taken in the core of Ursa Minor for additional photometric depth,
and a grid of 32 sets of frames were taken surrounding the core (see
Figure \ref{fig:grid}).  The objective of the observational program is
to detect giant stars in the top few magnitudes of the Ursa Minor giant
branch, and thus only short exposures were necessary.  For each core
field, exposures of 30, 30, and 300 seconds were taken in the
Washington $M$, Washington $T_2$, and $DDO51$ filters.  The grid fields
were observed with exposure times of 15, 15, and 150 seconds.  All
frames were reduced using the MSCRED Mosaic reduction package in
IRAF\footnote[2] {IRAF is distributed by the National Optical Astronomy
Observatories, which are operated by the Association of Universities
for Research in Astronomy, Inc., under cooperative agreement with the
National Science Foundation.}

Instrumental magnitudes for stars detected in all three filters were
derived using standard aperture photometry techniques.  Ursa Minor is a
very low surface density galaxy at high Galactic latitude, so at the
relatively shallow depth of our frames, even in the core regions only
$\sim 400-800$ objects were detected on each chip of the Mosaic
camera.  Thus, commonly used techniques of crowded field photometry
were not necessary.   For each chip of every Mosaic frame,
radial profiles were measured for many stars, and the average FWHM for
these stars was calculated.  Using the IRAF aperture photometry package
APPHOT, instrumental magnitudes were calculated for every object
detected on the chip using an aperture radius 2.5 times the average
stellar FWHM for that chip.  The average seeing during the observing
run was $\gtrsim1\arcsec$ so the average aperture radius was
$\sim10-11$ pixels.  Given an aperture of 11 pixels in radius, 800
stars cover only 0.3\% of the area of the chip.  In the event that an
aperture did include light from a nearby star, this object would likely
be rejected from the catalogue using the ``structural parameter'' described
in more detail below.

The instrumental aperture magnitudes for all detected objects were
transformed to the standard system defined by \citet{geisler90, geisler96}
standard stars.  Instrumental magnitudes for the standard stars were
calculated using a process identical to the one used for the Ursa Minor
survey region stars.  Photometric transformation equations of the
form:  
\begin{equation} 
m - M = k_{1} + k_{2}(X) + k_{3}(M-T_2)
\end{equation} 
\noindent were derived that included nightly zero point terms ($k_1$),
airmass ($k_2$), and color terms ($k_3$; except in the case of $T_2$,
which required no color term) from the instrumental magnitudes and
colors of the standard stars.  For the $M$ filter, the $M - T_2$ color
was used for deriving the transformation coefficients, and for $DDO51$,
the $M - DDO51$ color was used. The equations were derived using
software written by C.~P.\ that implements the matrix inversion
algorithm of \citet{harris81}.

After the photometric transformations were applied to the program
stars, the calibrated magnitudes of stars measured multiple times
(those found in the overlapping areas of each chip) were compared for
the purpose of refining the calibration.  Although the observations
were taken under photometric conditions, the magnitudes of stars
measured multiple times have dispersions of 
$0.006 - 0.009$ magnitudes, indicating that small errors may remain in
the photometry due to small changes in the transparency of the
atmosphere or possibly due to the limited accuracy of the airmass
coefficient and the airmass measurements.  An average offset for each
frame was calculated using the measurements of the overlapping stars
\citep[for a description of the technique, see][]{siegel2001}, and
these offsets (typically $< 0.003$ magnitudes) were applied to the
calibrated magnitudes, placing the photometry onto a
global system with a relative precision of 0.001 magnitudes.

\begin{figure}
\plotone{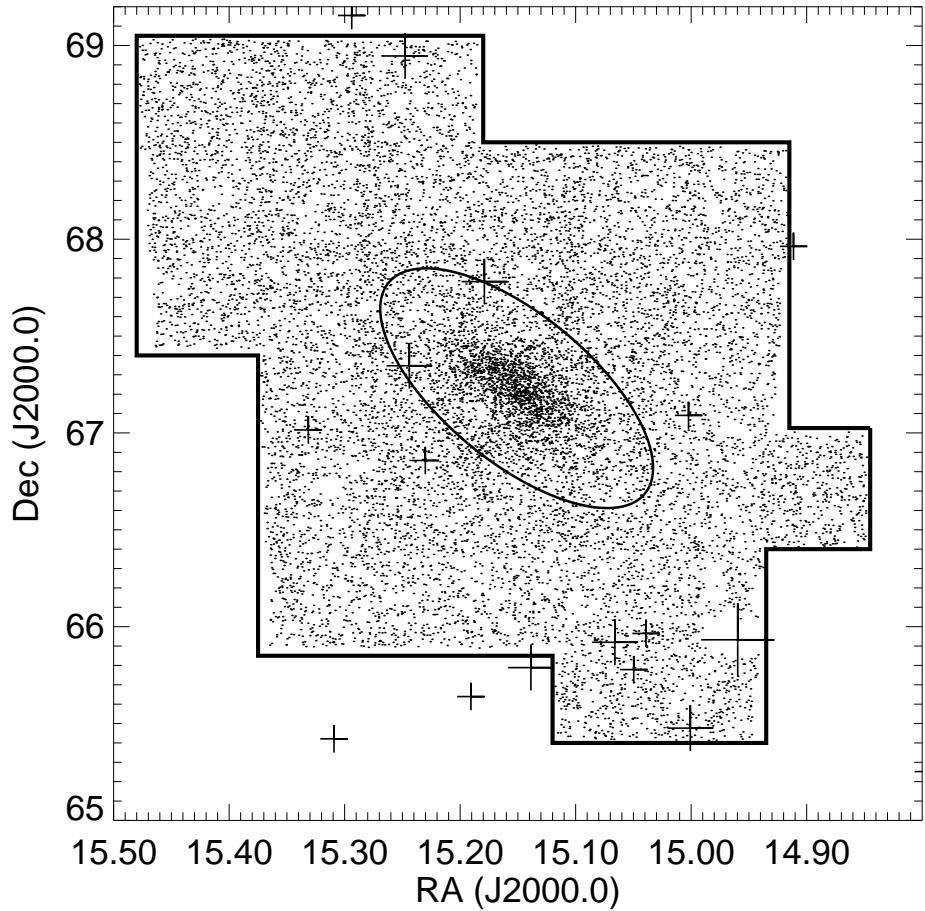}
\caption{Map of all stars detected in the survey region, which is
centered on Ursa Minor.  All fields observed during this program were
taken under photometric conditions.  The solid line gives a rough
indication of the boundaries of the survey.  The ellipse represents the
shape of Ursa Minor derived from the shallow data of K98.
The semi-major axis is set at the $\sim51\arcmin$ tidal radius
(IH95), which may or may not agree with the value derived by
K98 (see \S4.2).  The crosses mark the locations of
the brightest stars found in our survey region.  The largest cross
marks the location of the $V=4.7$ magnitude variable star RR UMi, which
saturated almost an entire chip of the Mosaic camera.  The slightly smaller
crosses mark the locations of stars with $5 < V < 7$, while the smallest
crosses mark the locations of stars with $7 < V < 8$.  The combination
of the bright stars along the southern edge of the survey and  bright
moonlight created a gradient in the limiting magnitude of the survey
across the grid.}
\label{fig:allstars}
\end{figure}

Non-stellar and other problem objects (e.g., close double stars) were
removed from the catalogue using a rough structural parameter derived
from the aperture photometry.  A second magnitude was calculated for
every object using an aperture 0.5 times the average FWHM of the stars
on the frame.  The difference between this magnitude and the 2.5 FWHM
aperture magnitude yields an estimate of the concentration of the light
from each object, which should be a fixed ratio for stellar images on
one frame.  Since the scatter around the mean value of this
``concentration parameter'' increases as the photometric error
increases, a running mean and standard deviation of the concentration
parameter as a function of magnitude was calculated, and all outliers
from the stellar locus were thrown out.  Finally, all stars with
measurement errors greater than 0.1 in any of the three filters were
also removed from the catalogue.  Astrometry for all stars was
determined by solving for the plate coefficients of each Mosaic chip
after matching $\sim$50 bright stars per chip with their positions in
the USNO-A2.0 catalogue \citep{usno}.   Figure \ref{fig:allstars}
displays the spatial distribution (in the USNO-A2.0 astrometric system,
J2000.0) of all stars detected in the survey that were retained after
the structural parameter and error cuts were applied to the original
catalogue.

\begin{figure}
\plotone{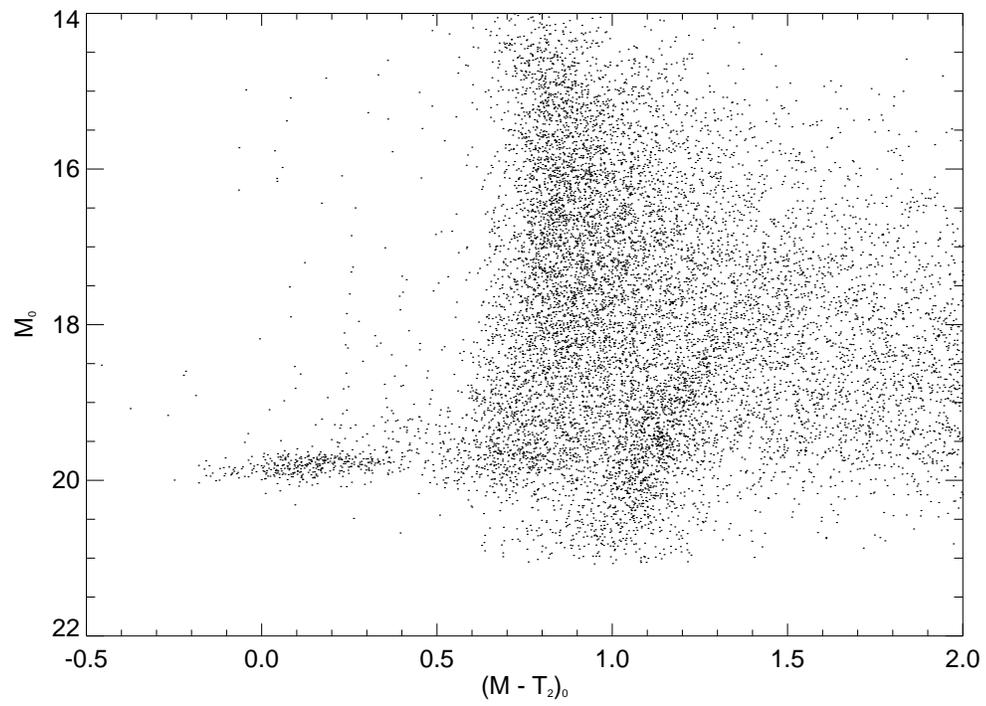}
\caption{Dereddened ($M - T_2$, $M$)$_0$ color-magnitude diagram for all stellar
objects found in our survey area.  Only objects with stellar profiles and with
magnitude errors $< 0.1$ in all three filters are included.}
\label{fig:allcmd}
\end{figure}

\begin{figure}
\plotone{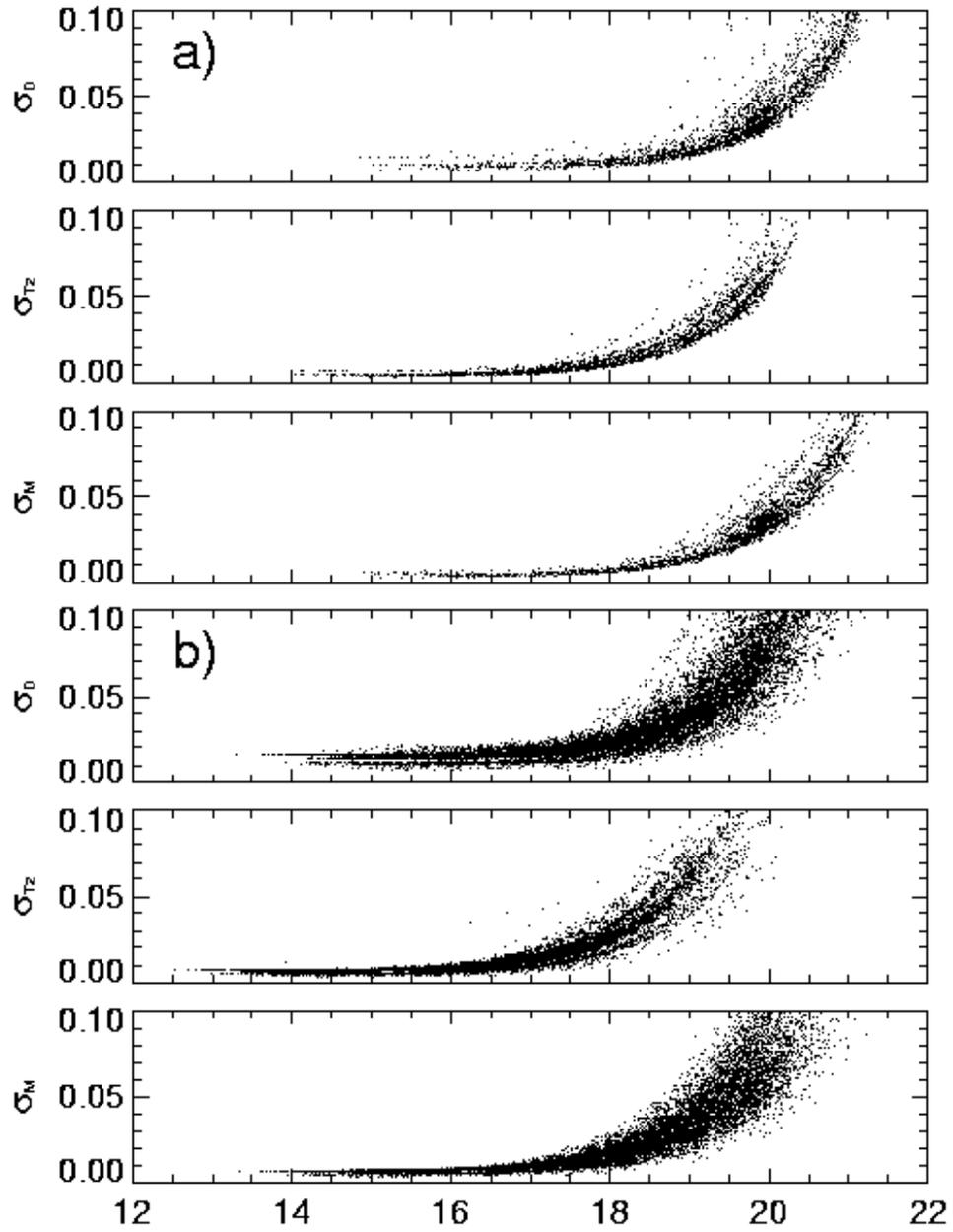}
\caption{Photometric errors for stellar objects in the survey region as a
function of magnitude for the (a) core fields and (b) surrounding fields.}
\label{fig:errs}
\end{figure}

An ($M - T_2$, $M$)$_0$ color-magnitude diagram (CMD) for all of the
stars from Figure \ref{fig:allstars} is presented in Figure
\ref{fig:allcmd}.  Each star shown in the CMD has been corrected for
reddening based on its celestial coordinates and a comparison to the
\citet{schlegel} reddening maps.  Since the core fields were observed
with exposure times twice as long as the grid fields and also because
the observations were taken with varying amounts of moonlight, the
limiting magnitude varies across our survey region from $M_0 < 19.4$ to
$M_0 < 21.0$.  This variation is apparent in the faint end of the CMD,
where the core RGB extends to $M_0 \sim 21$.  The core fields go deep
enough that the blue horizontal branch (BHB) stars are well detected at
$M_0 \sim 19.75$.  BHB stars are also detected in the surrounding grid
fields, but with larger photometric errors.  A plot of the photometric
errors as a function of magnitude is presented in Figure
\ref{fig:errs}.

\section{Selection of Candidate Ursa Minor Stars}

We use the technique described in \citet{haloI} to select giant stars
in Ursa Minor and to eliminate the majority of possible sample
contaminants.  Candidate Ursa Minor giant stars are selected to fulfill
two criteria:  (1) The stars must have magnesium line/band strengths
consistent with those of giant stars, and (2) the stars must have
effective temperatures and apparent magnitudes that place them within
the giant branch locus of known Ursa Minor giants in the
color-magnitude diagram.  Only those stars that meet criterion 1 are
tested with criterion 2, and only those stars that meet both criteria
are selected as candidate Ursa Minor giants.

Table \ref{photom} lists the positions and photometry for all candidate
UMi RGB stars (\S3.2) and UMi BHB stars (\S3.6) discussed below.

\subsection{Brief Description of the Photometric Selection of Giant Stars}

Here is a brief summary of the giant selection technique described in
more detail by \citet{haloI}:  Each field is observed in three filters,
Washington $M$, Washington $T_2$, and $DDO51$.  The $DDO51$ filter is
an intermediate band filter centered at 5150 \AA\ that measures the
strength of the MgH$+$Mgb triplet feature in the stellar spectrum.
This particular spectral feature is dependent on surface gravity in
late type stars, and is thus a good discriminator between K giant stars
and K dwarf stars.  \citet{geisler84} proposed using the Washington $M$
filter to measure the nearby stellar continuum so that the color $(M -
DDO51)$ can be used for luminosity classification.  While the $(M -
DDO51)$ color is primarily sensitive to surface gravity, since more
metal-poor giants would be expected to have less Mg absorption than
stars of higher metallicity, the index is secondarily sensitive to
metal abundance.  The equivalent width of the Mg feature depends on
stellar effective temperature, as well as gravity and abundance.  Since
we are specifically interested in K giants (because the sensitivity of
the Mg feature to gravity is strongest for this spectral type), we must
measure $T_{eff}$, also.  The $(M - T_2)$ color is very similar to the
standard $(V - I)$ color \citep{haloI}, which is a useful measure of
stellar effective temperature for late type stars.  Therefore, an
$(M - T_2, M - DDO51)$ two-color diagram can be thought of as an 
$(T_{eff}, \textrm{gravity})$ diagram useful for isolating K giant stars
from K dwarf stars.

\begin{figure}
\plotone{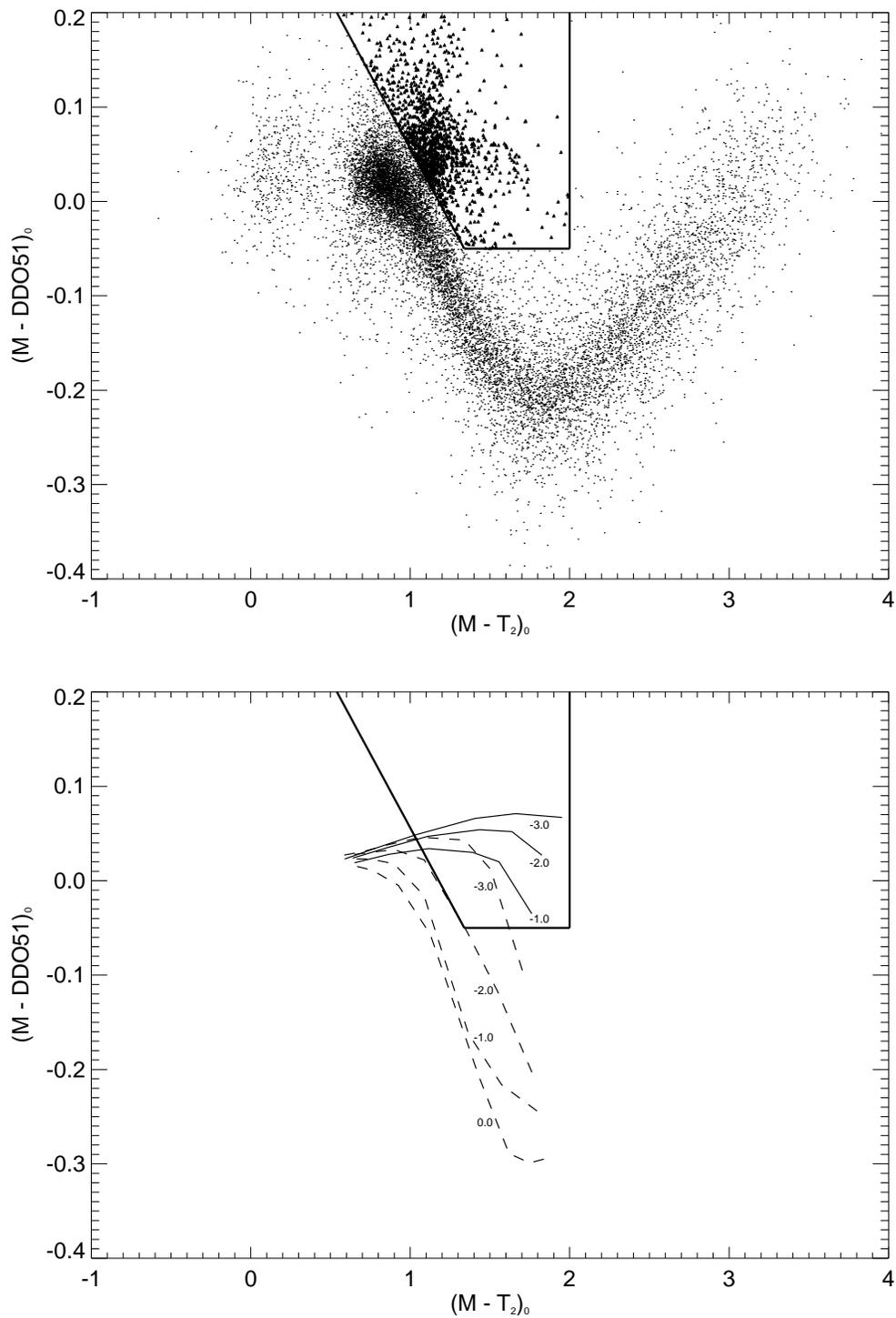}
\caption{The top panel is the ($M - T_2$, $M - DDO51$)$_0$ color-color
diagram for stars shown in Figure \ref{fig:allcmd}.  Dwarf stars lie
along the prominent, elbow-shaped locus in the diagram.  Giant stars
(plotted as filled triangles) lie predominantly in the region bounded by the
solid line seen in both panels.  The expected isochrones for dwarfs and
giants of specific metallicities \citep[derived from synthetic spectra
of][]{pb94} are shown as dashed and solid lines respectively in the
lower panel.}
\label{fig:allccd}
\end{figure}

The top panel of Figure \ref{fig:allccd} is the dereddened two-color
diagram (2CD) for all of the stars that were not eliminated by our
photometric error or structural parameter cut.  Dwarf stars lie along
the prominent, elbow-shaped locus, due to their strong magnesium
absorption.  The giant region, bounded approximately by the solid line
in Figure \ref{fig:allccd}, selects giant stars more metal-poor
than $[\rm{Fe}/\rm{H}] \sim -0.5$ \citep{haloI}.  For reference, the
lower panel of the Figure presents the expected isochrones of giant
stars and dwarf stars of various metallicities, which were 
modified from the synthetic spectra of \citet{pb94} \citep[see][for a
discussion]{haloI}.  The metallicity of Ursa Minor is assumed to be
$[\rm{Fe}/\rm{H}] \sim -2.2$ based on comparison of its RGB to that of
the globular cluster M92 \citep[e.g.,][]{mb99}.  However, high
resolution spectra of a sample of Ursa Minor giants suggest there is a
0.73 dex spread in the metallicity of Ursa Minor stars with an average
of $[\rm{Fe}/\rm{H}] = -1.90$ \citep{shetrone01}.  In any case, at
these abundances, the giants in Ursa Minor are expected to be well
separated from disk dwarfs in the 2CD.

Stars are considered to be giant candidates if they lie within the
bounded giant region in color-color space.  The boundary is drawn such
that the blue edge is approximately parallel to the dwarf locus, but
offset enough so that photometric error will not scatter too many
dwarfs into the giant region.  Of the 14,100 stars in Figure
\ref{fig:allcmd}, 1,342 of them are selected as giant candidates using
the selection box in Figure \ref{fig:allccd}.  The two-color selection
process is not perfect at selecting Ursa Minor giants; photometric
error can scatter solar metallicity dwarfs into the selection box, and
the intrinsic properties of subdwarfs and field giants will place them
in the selection box, too.  Our calculations \citep{haloIII} suggest
that the level of contamination of the sample by solar metallicity dwarfs
is likely to be $< 20$\%.

\subsection{Selection of Ursa Minor Giant Candidates}

Since the color-color diagram does not easily separate Ursa Minor
giants from the several types of potential contaminants mentioned  
previously, we rely on a second criterion to reduce
contamination and to select a more pure sample of Ursa Minor giants.
Because the RGB of Ursa Minor appears prominently in the ($M - T_2$,
$M$)$_0$ color-magnitude diagram, we can eliminate a large number of 
contaminant stars selected with the color-color criterion by
designating only those giant candidates that also lie along the Ursa
Minor RGB in color-magnitude space as Ursa Minor giant candidates.  In
order to delineate the RGB locus accurately, we have matched stars in
our catalogue to those in the proper motion catalogue of
\citet{kmc01u}.  Figure \ref{fig:kmccmd} shows the ($M - T_2$, $M$)$_0$
color-magnitude diagram for those stars in our catalogue that have
proper motion membership probabilities (i.e., their individual proper
motions are similar to the mean motion of the galaxy) $ > 75$\%.  The
``box'' enclosing the proper motion selected Ursa Minor giants defines our
second selection criterion; only those color-color selected giants that
{\em also} fall in the color-magnitude RGB box are considered Ursa
Minor giant candidates. Applying both photometric selection criteria,
color-color and color-magnitude, to our entire sample of 14,100 stars,
we have culled 788 candidate Ursa Minor giant stars (see Table
\ref{photom}).  A CMD
highlighting the selected candidate Ursa Minor giants as well as a plot
of their spatial distribution is presented as Figure \ref{fig:umingi}.
We note that at this point of the
analysis, 202 of the candidate Ursa Minor giants lie {\em outside}
the IH95 tidal radius of Ursa Minor.  We discuss the selection
of Ursa Minor BHB star candidates in \S3.6.

\subsection{Effect of Limiting Magnitude Variations}

Due to a combination of the presence of several nearby, very bright
stars (see Figure \ref{fig:allstars}) and significant moonlight during
the observations (the Moon was nearly full), there is a large variation
in the limiting magnitude of the various grid fields.  Also, three sets
of overlapping images of the Ursa Minor core (the RGB box in Figure
\ref{fig:kmccmd} was defined using stars found only in this region)
were taken with exposure times twice as long as those used for the grid
fields (this was done so that we could accurately define the locus of
the Ursa Minor giant branch in the region with the highest density of
member stars).  Therefore, the sample of Ursa Minor candidate giant
stars presented in Figure \ref{fig:umingi} is incomplete at the faint
end.

In order to accommodate the varying limiting magnitudes and to reflect more
realistically the density distribution of Ursa Minor
giant candidates, we analyze three separate subsamples of these stars
with magnitude limits of $M_0 = 19.3$, 19.65, and 20.0.  The entire
survey area (9.06 square degrees) is included in the $M_0 \leq 19.3$
sample.  In the $M_0 \leq 19.65$ and $M_0 \leq 20.0$ subsamples,
however, we do not include in our analysis those subfields that have
magnitude limits brighter than 19.65 or 20.0, respectively.  The $M_0
\leq 19.65$ subsample covers 8.31 square degrees, while the $M_0 \leq
20.0$ subsample covers 5.56 square degrees.  Figure \ref{fig:magcuts}
shows the Ursa Minor giant candidates and areal coverage for each of
these three magnitude-limited subsamples.

\subsection{Evaluation of Giant Background Level}

\begin{figure}
\plotone{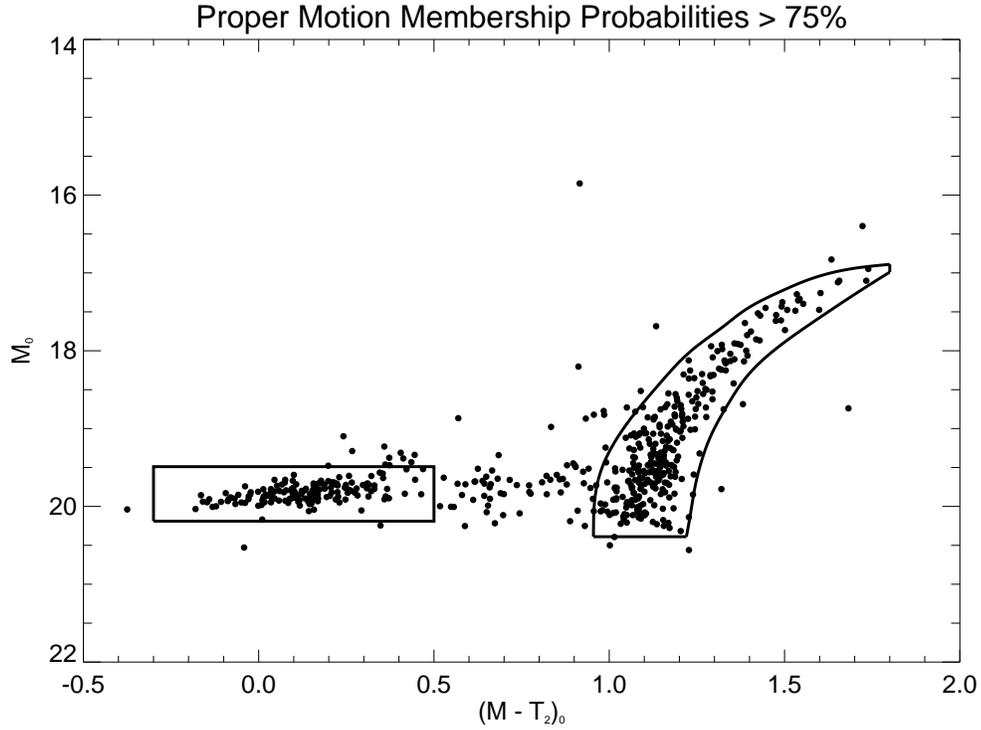}
\caption{($M - T_2$, $M$)$_0$ color-magnitude diagram for those stars
in our catalogue that also have proper motion membership probabilities $ > 75$\%
in the \citet{kmc01u} catalogue.  The adopted RGB locus for
Ursa Minor is designed to contain most of these proper motion-selected
giants while limiting contamination from field giants.  Blue Horizontal
Branch stars are also easily visible in this diagram, and we draw a
BHB bounding box useful for selecting these stars over the entire 
survey region, as well.  We discuss the BHB selection in \S3.6}
\label{fig:kmccmd}
\end{figure}

\begin{figure}
\plotone{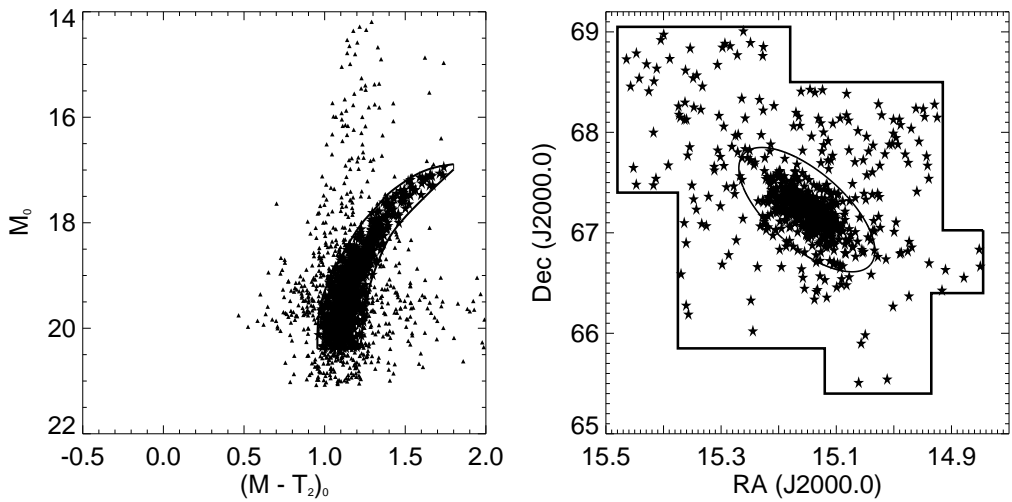}
\caption{Selection of candidate Ursa Minor giant stars.  The left panel is the
CMD for stars selected to be giants in our catalogue; triangles represent
those stars selected as giant stars with the color-color cut
from Figure \ref{fig:allccd}, while those giants that also lie in
the Ursa Minor RGB region of the CMD are plotted as stars.
The right panel shows the distribution of the Ursa Minor giant 
candidates on the sky.  As in Figure \ref{fig:allstars}, the
ellipse represents the previous measurements of Ursa Minor's shape and
tidal radius.  The relative paucity of stars at southern declinations
is a reflection of the variation in limiting magnitude across our 
survey area; the southwestern fields have the brightest limiting 
magnitudes, and thus we detect fewer giants in this region.}
\label{fig:umingi}
\end{figure}

Figure \ref{fig:umingi} demonstrates that there are field giants (i.e.,
those stars found to be giants in the color-color diagram that do not
lie within the UMi RGB box) in our survey area at a range of $M_0$
magnitudes.  Although we are likely eliminating the majority of dwarf
stars from our sample of candidate UMi giants by employing color-color
selection, we are unable {\em a priori} to remove from our sample those
field giants and field extreme subdwarfs that happen to have the combination of
distance, temperature, and abundance characteristics that place them
within the RGB bounding box in Figure \ref{fig:kmccmd}.  Thus, we must
evaluate the level of contamination expected from field giants and
field subdwarfs.

\citet{haloII} argue that the number of halo field giants per unit
solid angle should be flat to first order if the density of halo stars
is roughly an $R^{-3}$ power law.  Therefore, if we were to offset our
Ursa Minor RGB bounding box to brighter magnitudes\footnote[3]{In
principle, this should also work for offsets to fainter magnitudes, but
the incompleteness of our survey at the faint end will produce
incorrect results for the counts.}, the number of giants in the box
should remain roughly constant as a function of magnitude offset.
Indeed, for Carina, this was found to be the case \citep{haloII}.  We
repeat this exercise here:  We have taken the Ursa Minor RGB box
pictured in Figure \ref{fig:kmccmd} and offset it to brighter $M_0$
magnitudes in increments of 0.33 magnitudes.  We calculate the number
of color-color selected giants that are found in the CMD giant box as a
function of magnitude offset of the RGB box.  The calculation was
performed separately for each of the three magnitude-limited
subsamples, and in each case the RGB box was altered from its shape in
Figure \ref{fig:kmccmd}:  the lower limit of the RGB bounding box at
magnitude offset 0 was set equal to the magnitude limit of the sample
(i.e., 19.3, 19.65, or 20.0).

The results of this analysis are summarized in Table \ref{bkgrndtab}.
The data in the table illustrate the limitations of this technique for
estimating the background; at small magnitude offsets, the RGB bounding
box still contains a number of Ursa Minor giant stars.  At the largest
magnitude offset, the sample is incomplete because of saturation of
bright stars on the CCD chips.  Unfortunately, this leaves only a few
bins for estimating the background.  In Table \ref{bkgrndtab} lines
have been drawn to indicate the magnitude offset limits of those bins
used to derive the background density for each Ursa Minor subsample.

Nevertheless, we take the average of the several bins that do not contain
any Ursa Minor giants and that are not affected by saturation at the bright
end and determine a background giant density.  The error in the background
density was calculated assuming a Poissonian probability distribution,
that is $\sigma = \sqrt{N}$, where $N$ is the average number of counts
for the subsample.   A summary of the counts
and background levels are presented in Table \ref{subsamps}.  In each
subsample, the number of candidate Ursa Minor giant stars found outside
the tidal radius is $> 3.5$ times the
expected number of field giants.

We have further attempted to estimate the expected density of
background giants by duplicating our Ursa Minor giant candidate
selection on photometry of stars found in an ``off'' field.  Since the
``off'' field does not contain the Ursa Minor RGB, presumably when we
apply our Ursa Minor giant selection criteria to this field, we are
{\em only} detecting field giants.  \citet{jco01} are conducting a
photometric survey of the satellites of the Andromeda Galaxy, and they
have provided us with calibrated Washington $M$, Washington $T_2$, and
$DDO51$ magnitudes for a large sample of stars in the field of And I.
The And I giant branch is visible in these data, however, due to the
increased distance modulus of this galaxy compared to Ursa Minor (And I
lies $>800$ kpc from the Milky Way), the tip of the And I RGB is
fainter than $M_0 = 22$.  Thus, when we apply our Ursa Minor RGB
selection criteria to these data, we are only selecting Galactic halo
giants and subdwarfs.  In the 0.36 deg$^2$ And I survey region, we find
a background density of $2.8\pm2.8$ deg$^{-2}$, $8.3\pm4.8$ deg$^{-2}$,
and $13.9\pm6.2$ deg$^{-2}$ using our $M_0 \leq 19.3$, $M_0 \leq
19.65$, and $M_0 \leq 20.0$ RGB selection regions, respectively.

And I and Ursa Minor are at roughly the same galactic longitude, $l =
122$ and $l = 105$, respectively.  At these longitudes, giant stars
found at $M \sim 18-20$ (corresponding to distances of tens of kpc)
would be beyond the disk, so the faint giants in both of these fields
are presumably halo stars.  Therefore, even though And I is
significantly closer to the Galactic plane than is Ursa Minor ($b =
-24.9$ versus $b = 44.8$), as a first approximation, the density of
field giants in either of these fields should be similar since the halo
is to first order spherical (i.e., the distribution of halo stars is
independent of latitude).  Although the inner halo is found to be
flattened, the distribution of halo stars outside of roughly $R_{gc} >
8$ kpc is found to show no significant flattening
\citep[e.g.,][]{slz90, lh94, chiba00, siegel2001}.   Since the distant,
background giants in the And I field are unlikely to be part of the
flattened, inner halo, the background density in the And I field should
be comparable to the density of background giants expected in the Ursa
Minor field.  One caveat that must be considered is that the And I
field is much smaller than the Ursa Minor survey region. Thus the
background densities derived from this field are based on a small
number of stars, and therefore the quantization noise is large (e.g.,
the background density for the $M_0 \leq 19.3$ sample is based on 1
star, and therefore the density is $2.8\pm2.8$ deg$^{-2}$).
Nevertheless, despite their larger uncertainty, the densities derived
from the And I data are useful as a ``sanity check''.  The background
densities derived from the And I field for the two fainter subsamples
are larger than those estimated using the magnitude offset technique.
However, given the limitations of the And I data, these 1 to 1.5
$\sigma$ differences are not unexpected.

\citet{siegel2001} has derived models for the density laws of Galactic
stars in the thin disk, thick disk, and halo stellar populations based
on star count observations.  According to their best fit Galactic
model, the thick disk is not expected to contribute any stars to our
UMi RGB selection region, because even at the bright limit of the
selection region, a giant star will be $> 10$ kpc from the Sun.  At
this distance, the stars will be entirely halo stars.  However,
contrary to our simplifying assumption, the Galactic latitude of the
And I field does increase the density of halo stars in this region.
The \citet{siegel2001} models predict a density of 4 giants per square
degree to a depth of $M_0 \leq 20$ at the Galactic coordinates of Ursa
Minor ($l=105, b=45$), but the density in the And I region ($l=122,
b=-25$) is predicted to be 12 giants per square degree.  These
predictions match remarkably well with our measured values, which
suggests the methodologies used to derive observational estimates of
the background level are reasonable.

For further calculations, we adopt the background densities presented
in Table \ref{subsamps}.   It is reassuring that the background
densities estimated from the And I field are $< 1.5\sigma$ larger than
those estimated directly from the Ursa Minor survey region and both are
consistent with predictions from the Galactic model of
\citet{siegel2001}.

\subsection{Spectroscopically Verified Ursa Minor Stars}

The photometric survey of Ursa Minor that is presented here can be used
to provide candidates for spectroscopic followup; radial velocity and
metallicity information will allow us to determine if our candidates
are {\em bona fide} members of the Ursa Minor system.  Analytical calculations
of the expected contamination rate \citep{haloIII} as well as our experience with
photometric selection of giants in other stellar systems \citep[e.g.,
Carina,][And I \& II, Guhathakurta et al.\ 2001]{haloII, haloIII}
support our assumption that the level of contamination in our candidate
sample is likely to be low.  Here, we present spectroscopic
verification that the photometric selection of Ursa Minor stars is
efficient at removing dwarfs and field giants from the candidate
sample.

\citet{har94} obtained spectra of 60 candidate Ursa Minor stars.  Using
radial velocities, they confirmed that 45 are Ursa Minor
members\footnote{Hargreaves et al. list 46 confirmed Ursa Minor
stars in their Table 1, however the star identified as CUD267 is listed
twice.}, 14 are foreground dwarfs, and one is a halo K-giant.  Of these
60 stars, 59 are also included in our catalogue.  In Figure
\ref{fig:harstars} we present a 2CD and CMD with the spectroscopically
verified stars highlighted.  All of the stars found in the three deeper
pointings of the core region of our survey of Ursa Minor are included
in the figure for reference.  We note that one of the 45 Ursa Minor
members is a known carbon star \citep[CUD122; see the notes to Table 2
in][]{taft95}, and this star is found near the edge of our giant
selection region in color-color space.  Another of the member stars
passes our color-color giant selection, but lies outside of our Ursa
Minor RGB selection region in color-magnitude space.

\citet{taft95} have also obtained spectra of a sample of candidate Ursa
Minor stars.  Their catalogue contains a heterogeneously selected set
of stars, taken from several sources.  There is some overlap between
the \citet{har94} and \citet{taft95} samples.  After removing those
stars also found in the \citet{har94} catalogue, there are 48 new stars
with radial velocities consistent with Ursa Minor membership and 47
stars with disk-like radial velocities (i.e., contaminants) in the
\citet{taft95} catalogue.  In Figure \ref{fig:taftstars} we present the
2CD and CMD of the core pointings of Ursa Minor, with the
spectroscopically analyzed stars from the \citet{taft95} catalogue
highlighted.  \citet{taft95} include one star, N98 (this star is
enclosed in a large circle in Figure \ref{fig:taftstars}), in their
table of Ursa Minor members even though its velocity ($-298.7$ km
s$^{-1}$) is $\sim$50 km s$^{-1}$ from the mean velocity of the dSph.
They tentatively conclude that this star is a member of UMi, but our
photometry suggests that it is not a member.  The list of radial
velocity members also includes several known carbon stars (enclosed in
squares in Figure \ref{fig:taftstars}), which fail our photometric
giant selection criteria.  Finally, several other radial velocity
members fail our giant selection even though they are neither marginal
candidates, like N98, nor carbon stars.   We suspect that several of
these may be AGB stars, which our color-magnitude RGB box was designed
not to include.

Figures \ref{fig:harstars} and \ref{fig:taftstars} verify the
efficiency of our photometric selection technique.  All of the dwarfs
in the \citet{har94} catalogue lie outside of our giant selection
region.  The lone field giant in their sample lies inside our
color-color giant selection region, however it lies outside of our
adopted Ursa Minor RGB bounding box, and thus we too classify this star
as a field giant.  Of the 45 {\em bona fide} Ursa Minor stars found in
the \citet{har94} catalogue, we successfully identified 43 of these as
Ursa Minor giants, rejecting two (including the carbon star) that lie
just outside the edges of our color-color and color-magnitude selection
regions.  A similar result is seen for the stars in the \citet{taft95}
catalogue.  All 47 of their dwarfs lie outside of our giant selection
region in color-color space.  Of the 48 verified UMi member stars in
the \citet{taft95} catalogue, we successfully reidentified many of
them, however, seven failed either one or both of our selection
criteria.  The stars that we failed to reidentify as members include a
marginal candidate with a velocity significantly different than the
mean velocity of UMi and several known carbon stars.  We speculate that
the UMi member in the \citet{har94} catalogue and several of the UMi
members in the \citet{taft95} catalogue rejected by us may be AGB
stars.  We note that in a recent study of proper motion-selected stars
in Ursa Minor, \citet{schesk01} found several high probability member
stars with colors and magnitudes consistent with those of AGB stars.

Although our photometric selection criteria do not find UMi members
with perfect accuracy, the fraction of UMi giants that we miss
(``missed detections'') appears to be low and may even be 0\% if the
majority of these stars are AGB stars and not RGB stars.  On the other
hand, among this combined sample of 154 stars, we were 100\% efficient
in rejecting all 61 contaminants (no ``false detections'').  We
conclude from these spectroscopic data that we are (1) efficiently
minimizing dwarf contamination, and (2) slightly underestimating (i.e.,
being somewhat conservative in our selection of) the number of true
Ursa Minor giants.

These tests of our photometric selection of giants are encouraging,
however we note that the spectroscopically confirmed members of UMi
observed to date are found at the bright end of the giant branch.
Photometric error is on the average smaller for the bright stars than
it is for the faint stars in our sample, so we expect the contamination
rate among the bright end to be less than that at the faint end.
Several lines of evidence suggest that the overall contamination rate
among our giant candidates is low, however, the exact level of
contamination of the faint end of the sample remains to be verified.

\subsection{Blue Horizontal Branch Stars}

Astrometry of Ursa Minor stars \citep[][and Figure
\ref{fig:kmccmd}]{kmc86} shows that the blue stars found in the CMD of
the field containing Ursa Minor have high membership probabilities,
thus confirming that these are Ursa Minor blue horizontal branch (BHB)
stars.  The Ursa Minor BHB stars are prominent in the CMD of this part
of the sky because there are very few field stars in the magnitude
range of our survey that are as blue, $(M - T_2)_0 < 0.5$, as BHB
stars.  The paucity of field stars in this color range makes the BHB
stars an excellent tracer of the spatial distribution of Ursa Minor and
a check on the results from the RGB stars.

A BHB star selection region is defined in ($M - T_2$, $M$)$_0$
color-magnitude space such that it encloses the BHB stars with high
membership probabilities in the \citet{kmc01u} catalogue (see Figure
\ref{fig:kmccmd}).  The blue edge of the instability strip appears to
occur near $(M - T_2)_0 \sim 0.4$, but the BHB selection box extends to
$(M - T_2)_0 = 0.5$, so we will select some UMi RR Lyrae stars as well
as BHB stars.   The red edge of the selection box was not designed to
cleanly select BHB stars and to exclude red horizontal branch or
instability strip stars; instead, the selection box was designed to
remain conservatively blueward of the blue edge of the field star
population, which is found at $(M - T_2)_0 \sim 0.7$.  With this
conservative red limit, photometric error is unlikely to scatter field
stars into the BHB selection box, but we may expect some contamination
by UMi variable stars, which is acceptable.  It would be useful to be
able to analyze the distribution of all of UMi's stellar populations,
including its RR Lyrae stars, red horizontal branch stars, and AGB
stars.  However, only the RGB stars (which we can cleanly select using
two-color and color magnitude criteria) and BHB stars (which we can
cleanly select using only a color magnitude box) are easily separated
from the Galactic foreground with the data we have available.

When we apply the color-magnitude selection box for BHB stars to our
entire survey area, we find 505 candidate BHB stars.  In the images of
the center of Ursa Minor, the limiting magnitude is $M_0 < 21$, and
thus all of the BHB stars in the core are well measured, having typical
errors significantly less than our $\sigma < 0.1$ error cut.  However,
the surrounding grid fields do not go as deep, and in many cases the
BHB stars are found at the magnitude limit of a particular frame.  In
Figure \ref{fig:bhb} (left panel) we present the CMD of BHB candidate
stars found in our survey area that have an error in all three filters
less than 0.2 magnitudes (this error cut is still small enough that
scatter from the substantial field MSTO should not affect the sample).
The right panel of Figure \ref{fig:bhb} shows the distribution of these
candidates on the sky.  For reference, the stars in the left panel of
Figure \ref{fig:bhb} are shown with their error bars, most of which
fall short of the blue edge of the field star population.  We note that
a more conservative selection with a $\sigma \leq 0.1$ error cut reduces
the sample to 406 candidate BHB stars and limiting the 
red edge of the BHB box to $(M - T_2)_0 = 0.4$ selects 379 candidate
BHB stars, 25 of which are outside the $50.6\arcmin$ tidal radius of
UMi.  In Table \ref{photom} are listed the 406 candidate BHB stars taken
from the sample with the
more conservative error cut applied.

The correspondence between the spatial distribution of the candidate
BHB stars (Figure \ref{fig:bhb}) and that of the candidate RGB
stars (Figure \ref{fig:magcuts}) is perhaps the best evidence that
our sample of candidate giant stars is largely free from contamination.
The contamination rate of the candidate BHB star sample is expected
to be low, so the similarity between the BHB candidates and the RGB
candidates suggests that the contamination rates in these two
independently selected samples are comparable.  We can estimate
the number of potential contaminants in our BHB sample using a similar
technique to the one used to estimate the giant background level in 
\S3.4.  

We have taken the BHB selection region (which is 0.7 magnitudes thick
in $M_0$) and offset it to brighter magnitudes by 0.7, 1.4, and 2.1
magnitudes.  The number of stars enclosed in the BHB selection box
after these offsets are 59, 23, and 11 stars, respectively.  It is
likely that the first offset of 0.7 magnitudes is not large enough to
avoid UMi variable stars marginally brighter than the BHB population,
so 59 stars may be an overestimate for the number of non UMi members we
expect in our BHB selection region.  The two brighter boxes may not
accurately represent the number of faint, blue stars to be found at the
magnitude of the UMi horizontal branch.  However, if we adopt 59 as an
upper limit to the contamination and 11 as a lower limit, the
contamination rate of the BHB sample due to field stars should be 2 --
12 \%.  Although the total contamination rate in the sample of BHB
stars appears to be low, we note that the majority of the survey area
lies outside the tidal radius of UMi, and thus the fractional 
contamination rate in the extratidal region is likely to be 
higher than it is inside the tidal radius of UMi.

\section{The Two-Dimensional Distribution of Ursa Minor Stars}

Among the population of Galactic satellite dwarf galaxies, Ursa Minor
is often considered to be one of the primary candidates for tidal
disruption due to its elongated and possibly double-peaked morphology
(\citealt{oa85}, IH95, \citealt{demers95}), the shape of its surface
density profile (IH95), and its spatial and dynamical association
with the Magellanic stream of dwarf galaxies \citep{wekdem76, lb82,
lb295, srm96, palma02}.  The spatial distributions of different samples
of candidate Ursa Minor stars presented here (Figures \ref{fig:umingi},
\ref{fig:magcuts}, and \ref{fig:bhb}); all show a significant extended
population, which lends support to the tidal disruption hypothesis.
Alternatively, these UMi stars may be bound within the potential of an
extended dark matter halo \citep[cf.][]{burkdm97}.  In the following
sections, we analyze the morphology and surface density profile for
Ursa Minor derived from the samples of candidate stars discussed
earlier to test the predictions of the Galactic tidal interaction
scenario.

\subsection{Morphological Peculiarities}

The majority of the satellite galaxies of the Milky Way are dwarf
spheroidals; this nomenclature derives from their shapes, which are for
the most part spherical or ellipsoidal (e.g., IH95).  Since
the dSphs are not isolated systems, but are instead evolving in the
gravitational potential of the Milky Way (which varies as seen by a
dSph in a non-circular orbit), it is feasible that upon closer
examination, their current morphologies may reflect the effects of this
evolution.  The two dSphs at the extremes of morphology are Leo II, at
$R_{gc} \sim 200$ kpc, which is mostly spherical \citep{siegelaas}, and
Sagittarius, at $R_{gc} \sim 16$ kpc, which has tidal streams of stars
and clusters that encircle the Galaxy \citep{mmetal98, srm99, sdss,
dd00, ibcarb}.  Since Ursa Minor is the closest dSph to the Galactic
Center ($R_{gc} \sim 65$ kpc) after Sagittarius, its morphology might
be expected to be more similar to Sgr than Leo II.

The peculiar morphology of Ursa Minor is well documented; \citet{oa85}
were the first to propose that the core of the galaxy contains
substructure in the form of two clumps of stars.  Subsequently,
\citet{demers95} identified an off-center clump of 78 stars in their
study of a small region in the core.  IH95 confirmed the
presence of two clumps of stars in the core of Ursa Minor, separated by
an angular distance of $\sim15\arcmin$.  From observations of the
largest area prior to our survey, K98 detected
substructure in the shape of their isodensity contours of Ursa Minor.
However, the authors concluded that the secondary peak visible in their
contour plot (and also detected in previous surveys; \citealt{oa85}, IH95)
is not detected at a statistically significant level.  On the other
hand, relying on a proper motion selected sample of UMi stars that is
expected to be nearly free of contamination, \citet{schesk01} instead
find that the internal substructure in UMi \textit{is} statistically
significant.  The most recent observations of the core by \citet{bd99}
were made with the WFPC2 camera on the {\it Hubble Space Telescope}.
With the high resolution afforded by the {\it HST} and the depth of the
data ($m_{F606W} \leq 24$), \citet{bd99} resolved the density peak near
the center, and claim that the enhancement is due to a ring of stars
surrounding a low density void.

\begin{figure}
\plotone{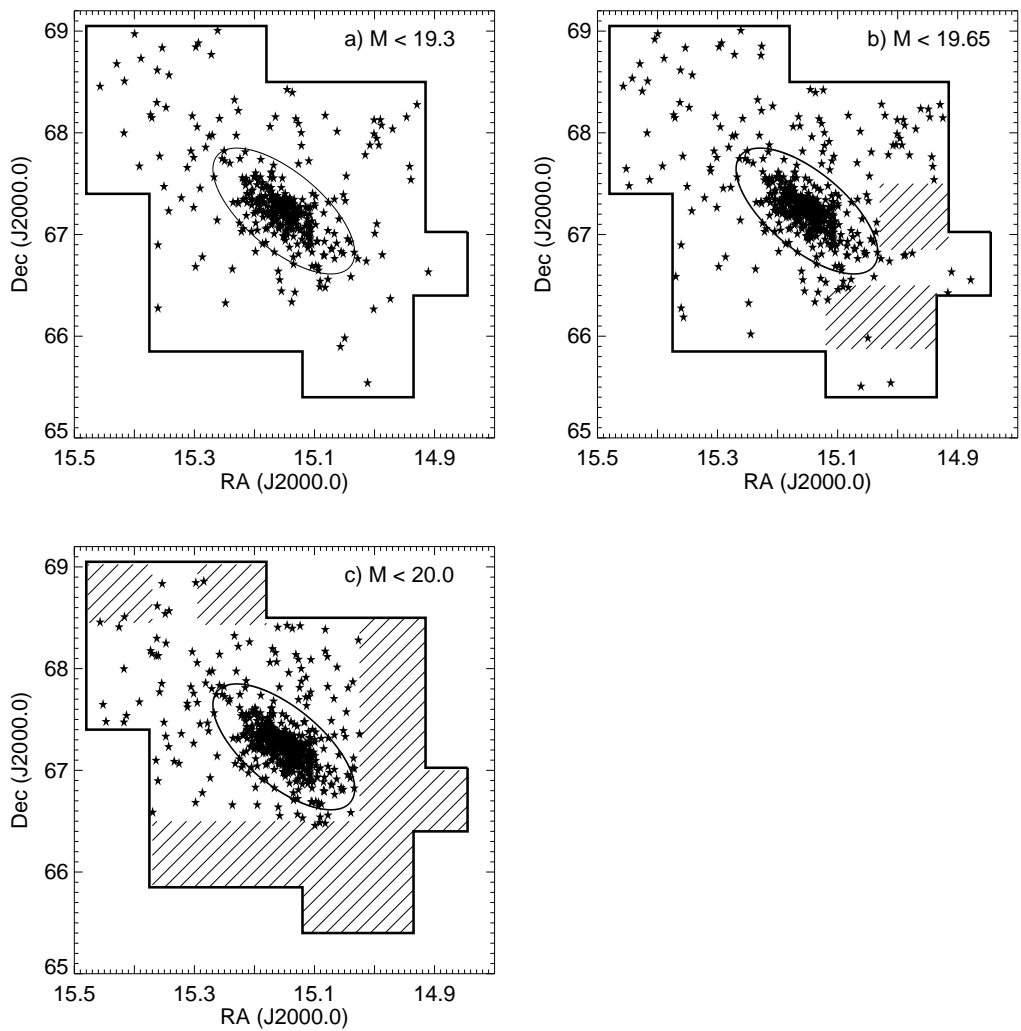}
\caption{Three magnitude limited subsamples of Ursa Minor giant
candidates.  Panel a (upper left) is the $M_0 \leq 19.3$ sample, panel
b (upper right) is the $M_0 \leq 19.65$ sample, and panel c (lower
left) is the $M_0 \leq 20.0$ sample.  The hatched areas in the two
fainter subsamples represent area excluded from the sample because
those frames were incomplete at the magnitude limit of the sample.  The
stars plotted within hatched regions in the panels were detected on
adjoining frames (all of our frames overlap by $5\arcmin$).}
\label{fig:magcuts}
\end{figure}

\begin{figure}
\plotone{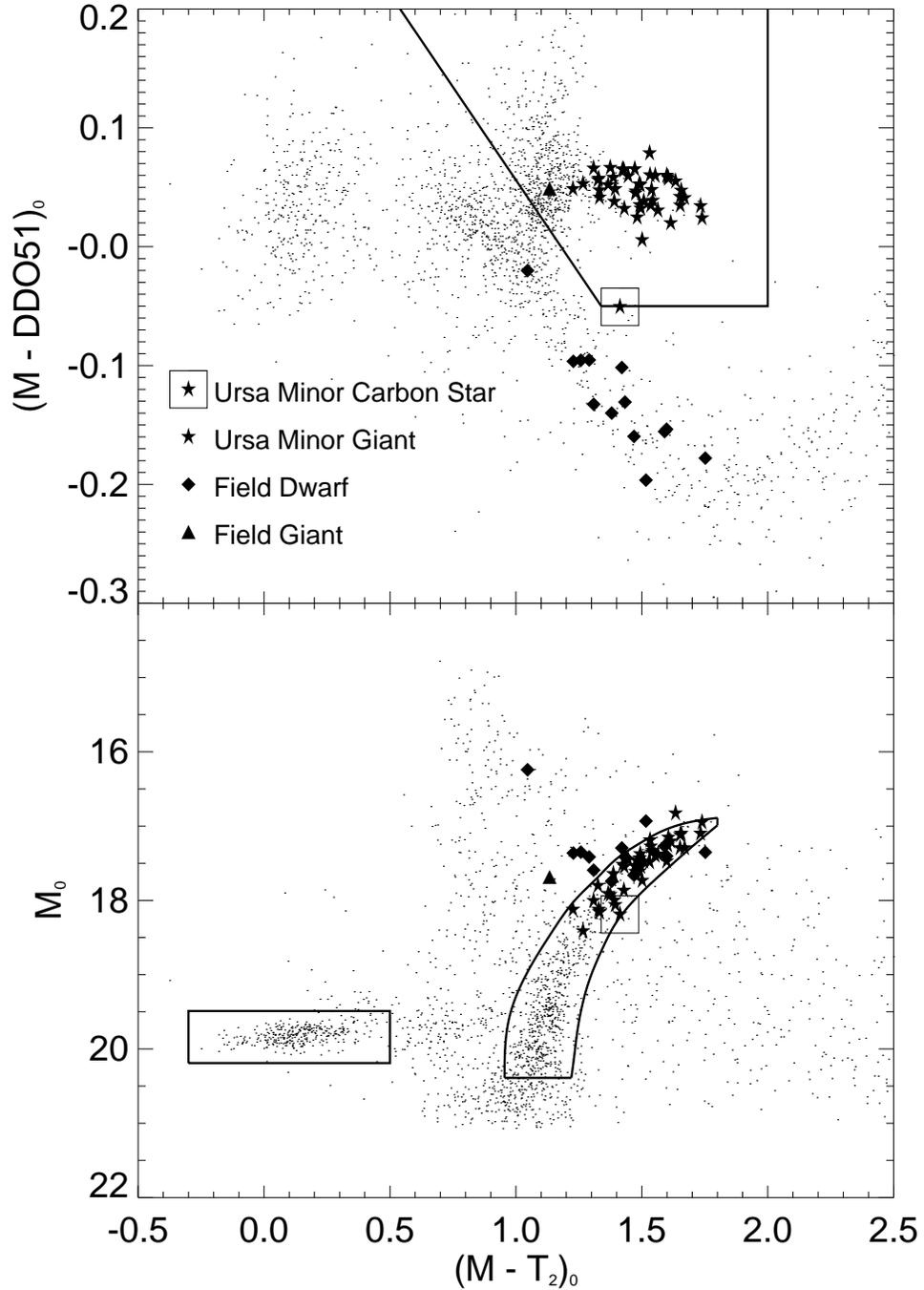}
\caption{The positions of spectroscopically verified Ursa Minor stars
and spectroscopically verified contaminants from the \citet{har94}
catalogue in color-color (upper panel) and color-magnitude (lower
panel) space. In the color-color diagram, the giants (filled triangle
and filled stars) and dwarfs (filled diamonds) separate cleanly.  All
of the dwarfs are found outside of our giant selection region, but the
halo K-giant is found inside of the giant region.  The carbon star in
the catalogue \citep[CUD122 in the catalogue of][]{har94} is found near
the edge of the giant selection region in color-color space.  All
except one \citep[EDO26 in the catalogue of][]{har94} of the
spectroscopically verified Ursa Minor giant candidates lie inside our
RGB bounding box in the CMD.  We speculate that this star may be an AGB
star or other type of post-RGB star.  The field halo K-giant lies well
outside of the CMD RGB box but was successfully classified as a giant.}
\label{fig:harstars}
\end{figure}

\begin{figure}
\plotone{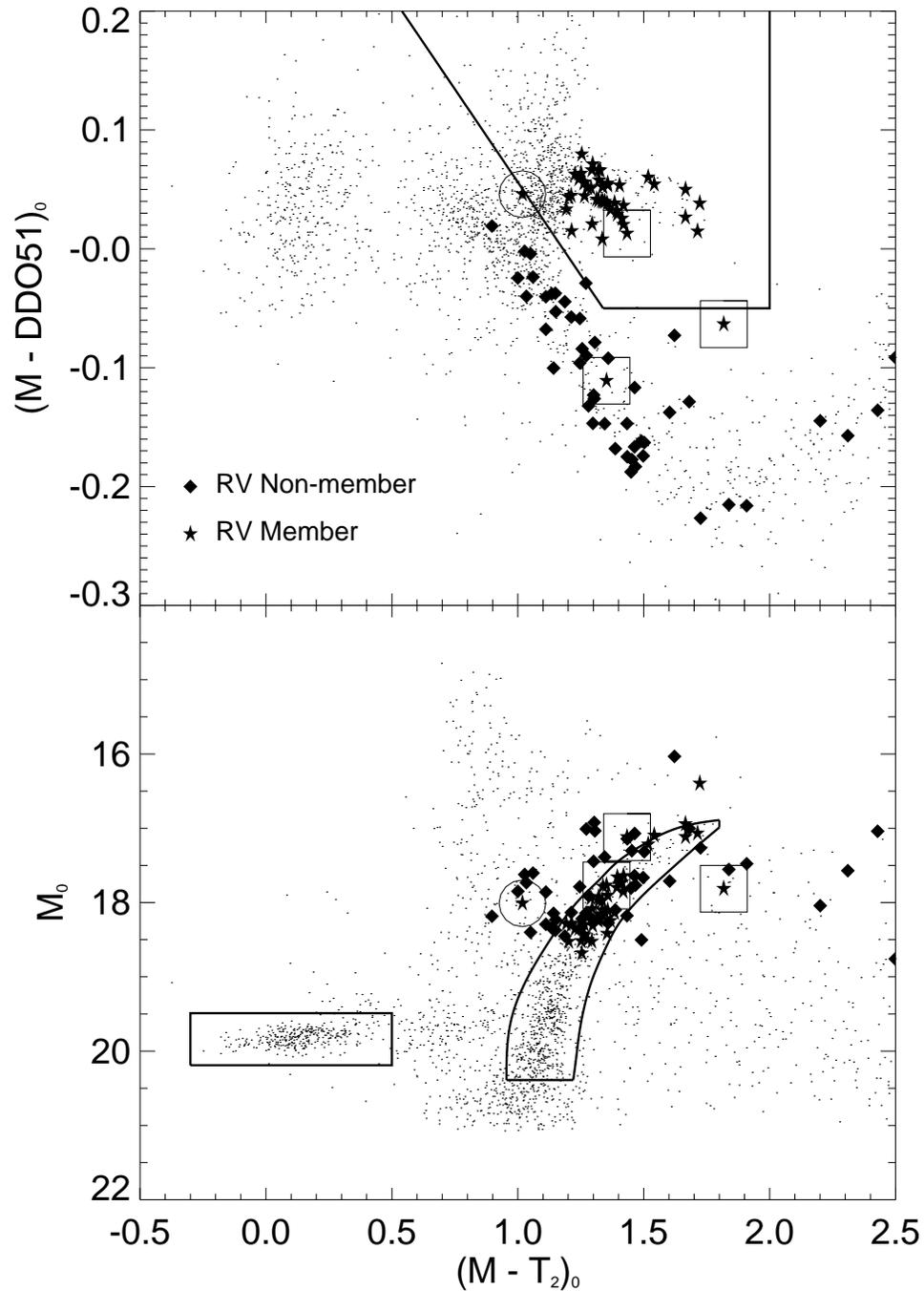}
\caption{The positions of spectroscopically verified Ursa Minor stars
and spectroscopically verified contaminants from the \citet{taft95}
catalogue (and not already shown in Figure \ref{fig:harstars}) in
color-color (upper panel) and color-magnitude (lower panel) space.  As
in Figure \ref{fig:harstars}, the contaminants present in the
\citet{taft95} catalogue appear to be entirely disk dwarfs, but in our
study, these stars separate cleanly from the color-color selected
giants in the color-color diagram.  Several of the stars identified as
Ursa Minor members by their radial velocity fail one or both of our UMi
giant selection criteria.  One of these stars, N98, is circled.  This
star has a velocity $\sim$50 km s$^{-1}$ from the systemic velocity of
UMi, and is only considered a possible member by \citet{taft95}.  Our
photometry suggests it is a non-member.  Several of the other radial
velocity members that fail our giant selection criteria are known
carbon stars; these three stars are enclosed in open squares.  The other
radial velocity members that fail our giant selection criteria may be
AGB stars, which our RGB selection region was not designed to include.}
\label{fig:taftstars}
\end{figure}

\begin{figure}
\plotone{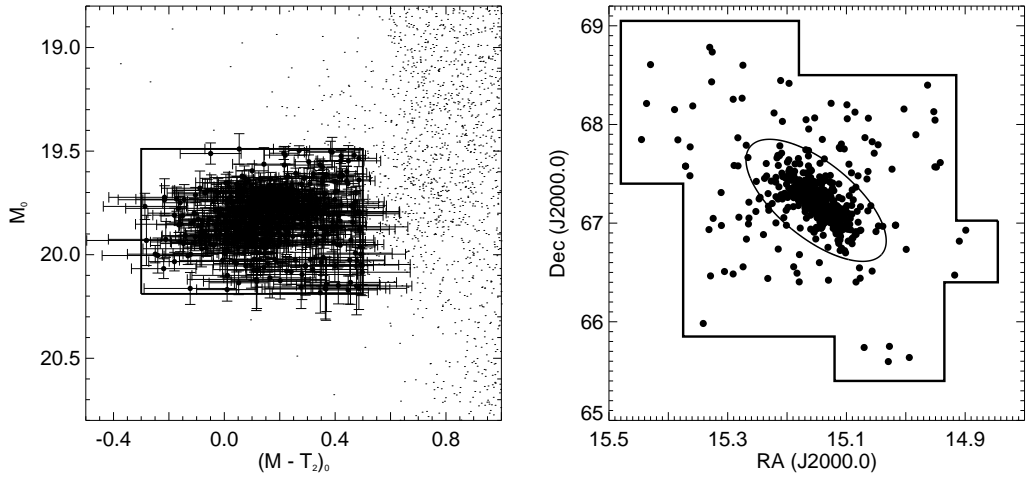}
\caption{Blue horizontal branch (BHB) stars in our Ursa Minor
catalogue.  The left panel is the ($M - T_2$, $M$)$_0$ CMD of the region
in color-magnitude space that includes Ursa Minor BHB stars.  The
bounding box used to select candidate BHB stars shown in Figure
\ref{fig:kmccmd} is reproduced here.  Also, each star is shown with its
error bars in both dimensions.  The right panel is the spatial
distribution of the Ursa Minor BHB candidates.  The region that appears
devoid of BHB stars is due to the variation of limiting magnitude among
the various grid fields; several grid fields do not go deep enough to
detect any BHB stars.}
\label{fig:bhb}
\end{figure}

\begin{figure}
\plotone{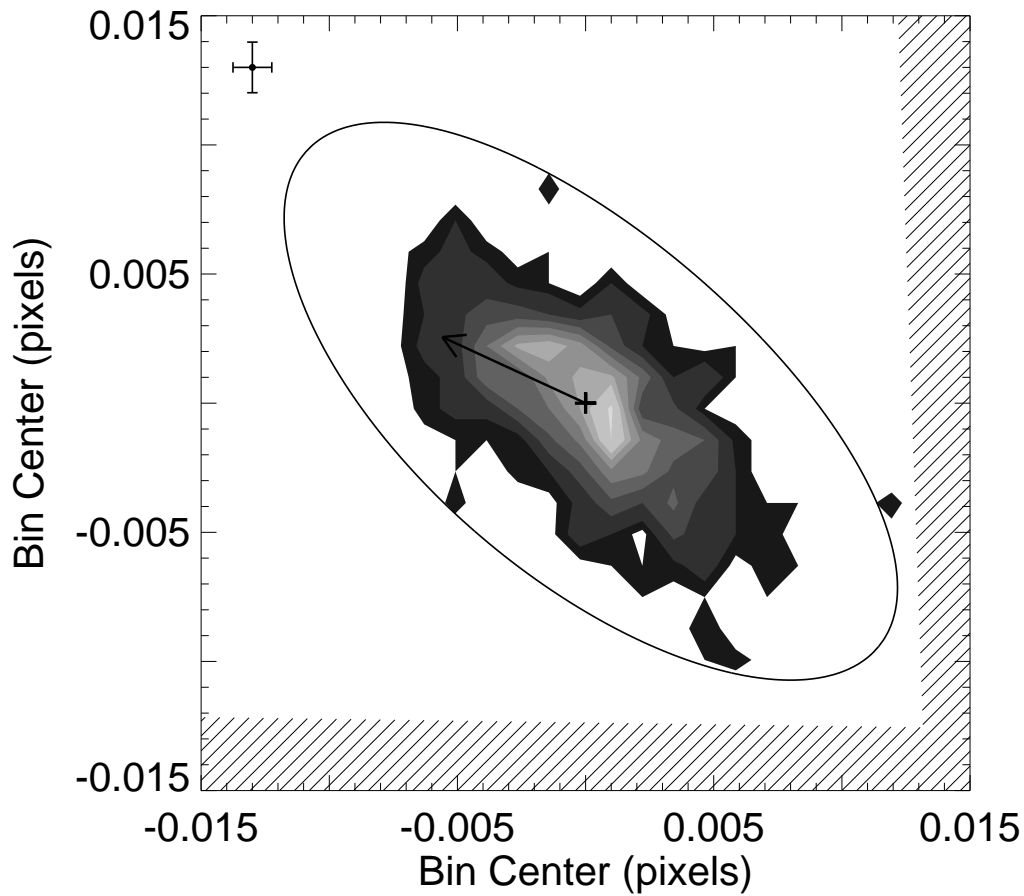}
\caption{Isodensity contours of Ursa Minor.  A sample of 1001 candidate
Ursa Minor RGB and BHB stars are represented by this image.  The
hatched region (reproduced from Figure \ref{fig:magcuts}) indicates
area within our survey region excluded from this analysis due to
incompleteness problems.  The equatorial coordinates of each RGB and
BHB star were converted to Cartesian using a tangential projection (the
image is presented such that north is up and east is to the left).  The
Cartesian space was divided into a grid of $50 \times 50$ ``pixels'' of
equal area ($\sim17.5$ square arcminutes), and stars were counted in
each grid cell.  This figure represents the inner $25 \times 25$ pixels
in the grid, and is 1.7 degrees on a side. The contour levels are
2,4,8,12,16,20,24,27,33,34 stars per pixel.  The adopted center
(K98) of Ursa Minor used in the projection is plotted as a
cross.  The direction of the orbit of UMi based on the \citet{kmc01u}
proper motion is indicated as an arrow.  The error bars plotted in the
upper left corner represent the uncertainty in the location of the end of
the arrow calculated from the proper motion.  Several
features are visible in the contours:  There do appear to be two
off-center peaks separated by a valley.  The secondary peak is
elongated in a direction between the major and minor axes.
Thus, the overall central morphology of Ursa Minor appears to be
crescent shaped or hooked, rather than elliptical.}
\label{fig:isodens}
\end{figure}

Figure \ref{fig:isodens} shows an isodensity contour plot for the
central region of Ursa Minor from our data.  The sample of candidate
Ursa Minor stars used to create this image is a combination of the $M_0
\leq 20.0$ RGB candidates from Figure \ref{fig:magcuts} (lower left
panel) and the BHB candidates with magnitude errors in each filter
$\leq 0.1$ mag (since the contour image includes almost entirely stars
found in the longer exposure time core fields, all of the BHB stars
have errors below this limit).  RGB and BHB candidates found in the
excluded fields in the lower left panel of Figure \ref{fig:magcuts} are
not included due to the incompleteness problems in these fields at the
magnitude of the HB.  Since the density of Ursa Minor stars in these
outer grid fields is low, this areal restriction does not change the
appearance of the isodensity contours (the hatched region that
indicates area excluded from the survey region for $M_0 \leq 20.0$ in
Figure \ref{fig:magcuts} is reproduced in Figure \ref{fig:isodens};
none of the contours lie within the excluded region).  The final sample
used contains 1001 stars.   The image was created using the following
steps:  (1)  The equatorial coordinates of each star were converted to
Cartesian coordinates using a tangential projection centered at the K98
center of Ursa Minor ($\alpha_{2000.0}$, $\delta_{2000.0} =$
15$^{\textrm{h}}$09$^{\textrm{m}}$03.9$^{\textrm{s}}$,
67\arcdeg13\arcmin51\arcsec).  (2) The tangential plane was partitioned
into a grid of $50 \times 50$ ``pixels'' each $4.2\arcmin$ on a side.
(3) The number of stars in each pixel were counted, creating a
two-dimensional array suitable for presentation as a contour plot.

Direct comparison between the contour plot presented here and previous
work is complicated by the difference in sample selection.  Previous
studies of Ursa Minor included many more stars than are presented here,
because these studies all probed deeper into the luminosity function of
Ursa Minor, reaching further down the giant branch (IH95, K98),
sometimes to the main sequence turn off \citep{oa85, bd99}.
However, in each of these studies, only single filter or dual filter
data was taken, and we would argue that the level of contamination by
non-Ursa Minor stars in these previous presentations of the isodensity
contours of Ursa Minor are less certain than our own work, which is
guided by the ability to discriminate between Galactic stars and Ursa
Minor stars with the use of Washington$+$DDO51 photometry.  Thus, we
expect the signal to noise in the outer isodensity contours presented
here to be better than previously available representations of Ursa
Minor, while the signal to noise in the core region is likely to be
similar.

The isodensity map of Ursa Minor shows several interesting features.
We do detect two off-center peaks in the spatial distribution: the
strongest just west of the center, and a secondary peak to the
northeast.  Within the IH95 core radius of $15.8\arcmin$, the
mean number of UMi candidate stars per pixel is $13.5$.  Due to the
small number of pixels within this area and the presence of the peaks
within the core region, the standard deviation in the number of UMi
stars per pixel within the core radius is $\sigma=9.8$.  Therefore, the
primary peak (34 counts) is a $2\sigma$ peak above the mean for the
core region, but the secondary peak (27 counts) is only a $1.5\sigma$
peak.  However, the stars that make up the primary peak are spread out
over two pixels of 34 and 33 stars each, so with slightly coarser
binning or with a slight change in center, this peak is found at much
higher significance.  The secondary peak is more diffuse, and remains
at $\sim1.5\sigma$ even if the bin size or phasing changes.  Thus, as
was found by K98 with their data, the secondary peak is
not detected with high statistical significance in our data, but both
surveys agree about the existence of this density enhancement at a
similar level.  Since several surveys have now identified this second
peak in UMi, albeit at low statistical significance in most cases, it
seems likely that the secondary density peak is a real feature in the
dSph, and the significance of it in any particular survey is limited by
the number of stars in Ursa Minor available at the magnitude limits
probed by most observers.  Another feature that is prominent in our
isodensity map that is seen to a lesser degree in IH95 is the
``hook'' in the northeast portion of Ursa Minor.  The stars that form
the secondary peak are elongated not along the major axis, but along a
line at a position angle between that of the major and minor axes.  The
bend in the contours is in the direction of the orbital motion of Ursa
Minor, which may indicate this clump of stars is being
stretched by the Galactic tidal field.

In an attempt to reduce the noise of the isodensity contours introduced
by the gridding process, we have smoothed the stellar positions using a
Gaussian kernel, creating the contour plot seen in Figure
\ref{fig:isodsmth}.  The following process was used for the smoothing:
(1) Each star was replaced by a two-dimensional square array
3.4$\arcmin$ on a side.  (2) The square array was filled with the
values calculated for a two-dimensional Gaussian with center at the
position of the star and a large FWHM, such that the values in the
square array were almost flat.  (3) The survey area was then
partitioned into a finer grid of $100 \times 100$ ``pixels'' each
$2.1\arcmin$ on a side, and the number of counts from each Gaussian
smoothed ``star'' in each pixel was calculated.

The smoothed representation of the stellar spatial
distribution reduces the significance of the secondary peak further,
while enhancing the primary peak, and reinforcing that the primary peak
is truly the location of the highest surface density of UMi stars.  The
primary peak is offset from the center of UMi as defined by the outer
contours, and it is also offset from the center of symmetry of the
``ring'' of UMi stars (plotted as a filled square) seen in the HST
images of Ursa Minor by \citet{bd99}.  The peak isodensity contours are
less elliptical than the outer contours, and their position angle is
different as well.  Overall, the contours appear to twist at increased
distance from the peak contours, giving the galaxy an overall
``\textsf{S}-shaped'' morphology.

In order to test the significance of the \textsf{S}-shaped morphology
of the UMi isodensity map, we have compared the distribution of UMi
stars within the elliptical boundary defined to have tidal
radius along the semi-major axis of 50.6$\arcmin$ to a model for a dSph 
with surface density distributed
according to the ellipticized single-component King model as
parameterized by K98.  Our model was created by selecting points
randomly within the elliptical boundary of UMi (drawn in Figure
\ref{fig:isodens}).  Using a rejection algorithm \citep[see \S7.3
in][]{numrec}, the random points were selected such that their 
surface density distribution follows that for the ellipticized King profile
of UMi.  Among the 1001 stars used to create Figure \ref{fig:isodens},
847 of them are within the elliptical boundary of UMi.  In order to 
increase the signal to noise of the model, it was generated with 10
times this number of stars, or 8470 points.  In order to test the 
null hypothesis that the UMi stars are drawn from an ellipticized
King model parent distribution, we compared the distribution of
the 847 UMi stars to the 8470 model stars with the two-dimensional 
Kolmogorov-Smirnov test \citep{ff87, numrec}.  The probability
that the 847 UMi stars are drawn from the same parent distribution
as the model is $<<1$\%.  Thus, we conclude that the distribution
of stars used to create Figure \ref{fig:isodens} deviates from a
symmetric, ellipsoidal model at a statisically significant level.

\subsection{Surface Density Profile}

For the purpose of measuring the important physical quantities such as
the mass or luminosity of a dSph, model fits are usually made to the
surface density profile of dSph stars.  It is generally assumed that
dSph galaxies can be fit with a King profile, however other models,
such as exponentials or S\'{e}rsic profiles, are used as well.  While it
can be argued which type of model is the most reasonable to use in
fitting a dSph profile, most previous studies of UMi have fit King profiles to
the observed stellar distribution (e.g., IH95, K98).  In this section,
we use the structural parameters for UMi derived from King profile fits
by IH95 and K98 to calculate a new surface density profile of UMi.  We
note here that there is an apparent discrepancy between the IH95 and
K98 fits to UMi that is relevant to the search for extratidal stars
associated with this dSph; the tidal radii derived from these two
studies differ significantly.  IH95 derived a tidal radius of $r_t =
50.6$\arcmin, while K98 quote a value of $r_t = 34.0$\arcmin.  However,
the former is a semi-major axis value, while the latter is a true
``radius'', that is, it is the tidal radius expected if Ursa Minor were
circular and not elliptical.  Converting the K98 value to a semi-major
axis value ($a_t = r_t \times \sqrt{1-\epsilon}$) gives 50.9\arcmin,
almost identical to the IH95 measurement.  For the purpose of
illustration of the structure of UMi, the position angle and
ellipticity parameters from the K98 study are adopted, however, we
adopt the semi-major axis value of $r_t = 50.6$\arcmin.  We adopted these
structural parameters because the K98 study is the largest area survey
prior to our own that probes the largest dynamic range of UMi stellar
density.  The ellipse seen in Figures \ref{fig:allstars},
\ref{fig:umingi}, \ref{fig:magcuts}, and \ref{fig:bhb} was constructed
using these shape parameters.

Using the standard method for estimating stellar densities in dSphs
(cf.\ IH95), we have constructed surface density profiles of the Ursa
Minor dSph using the three magnitude-limited subsamples of UMi RGB
candidates seen in Figure \ref{fig:magcuts}.  Due to the more serious
completeness problems in the sample of BHB candidates (several of the
survey fields don't go deep enough or blue enough to detect UMi BHB
stars at all) no profile was created using these stars.  However, by
visual inspection alone the distribution of BHB stars (Figure
\ref{fig:bhb}) appears quite similar to that of the RGB stars (Figure
\ref{fig:magcuts}). For the RGB stars, the densities were calculated by
first counting our candidate Ursa Minor giant stars in elliptical
annuli of successively larger semi-major axis.  The shape of the annuli
correspond to the structural parameters mentioned previously (derived
from the shallow data for Ursa Minor by K98; specifically, an
ellipticity of 0.554, a position angle of 49.4$^{\circ}$, and a center
of B1950.0 15$^{h}$08$^{m}$27.5$^{s}$, 
+67\arcdeg25\arcmin12\arcsec\ were adopted).  The center of Ursa
Minor is not well-defined, however, the galaxy is diffuse at low
densities, and small changes in the location of the center do not
affect significantly the number of stars in a given annulus.  The star
counts in each annulus were converted to densities (arcmin$^{-2}$) by
dividing by the area of the annulus.  The background was removed by
subtracting off the mean background density for each subsample (Table
\ref{subsamps}) from the Ursa Minor density calculated for each
annulus.  Within the core radius of IH95, we space our annuli in
intervals of $3.4\arcmin$, but in order to improve our signal to noise,
outside a radius of $13.6\arcmin$, the annuli are spaced at
$6.8\arcmin$.  The inner and outer semi-major axes of the annuli, the
area of each annulus, and the raw Ursa Minor RGB counts are given in
Table \ref{surfdens}. For the $M_0 \leq 19.3$ UMi RGB candidates, the
outermost annulus that fits completely within the boundaries of the
survey region has $a = 95.2\arcmin$. The survey fields not included in
the fainter samples due to photometric incompleteness restrict the size
of the largest annulus that fits completely within the boundary region
to $61.2\arcmin$ and $54.4\arcmin$ for the $M_0 \leq 19.65$ and $M_0
\leq 20.0$ samples, respectively.  We made star counts in annuli with
outer radii up to $204.0\arcmin$, and we determined numerically the
fractional area of each annulus enclosed within our survey region.
Thus, Table \ref{surfdens} includes star counts for each subsample out
to $r = 204.0\arcmin$, however, the increasingly reduced fractional
area in the outermost annuli introduces increasingly more noise into
those bins.  For this reason, the outermost annuli ($r_{outer} >
95.2\arcmin$) have been spaced at $27.2\arcmin$.

Figure \ref{fig:profile} presents the radial surface density profile
derived for Ursa Minor from the star counts in Table \ref{surfdens}.
Shown in the upper panel are the results for the three
magnitude-limited subsamples of Ursa Minor giant candidates:  the $M_0
\leq 19.3$ sample (filled circles), the $M_0 \leq 19.65$ sample (filled
triangles), and the $M_0 \leq 20.0$ sample (filled diamonds).  In both
panels, each set of points was offset vertically by one order of
magnitude (i.e., one tick mark along the logarithmic y-axis) in order
to reduce confusion due to overlapping points.  In the upper panel,
those points that are derived from annuli that do not fit completely
within the boundaries of our survey region (and thus may be susceptible
to local density fluctuations since they are not completely sampled)
are represented as open symbols.   In addition, the deeper, background
corrected Ursa Minor star counts of IH95 are shown (filled
stars), for comparison.  After normalizing these four profiles so that
their densities at $R \sim 6.8\arcmin$ are all equal to 1.0 (this point
was chosen since it is at high signal to noise and the scatter at this
radius between our three subsamples is low), they can be more easily
compared (Figure \ref{fig:profile}, lower panel).  The IH95 
King profile fit has tidal radius $r_t = 50.6\arcmin$; this
fit is plotted as a solid line in the lower panel of Figure
\ref{fig:profile}.  Although this fit follows the IH95 points
well, our starcounts appear to deviate away from the fit; the densities
for each magnitude limited sample measured in this study are found to
be systematically larger than the fit for all radii $r > 20.4\arcmin$.
The King profile derived by IH95 is an especially poor fit to
the bright sample of Ursa Minor giant candidates; past $20.4\arcmin$,
these points are fit very well by a power law $r^{-\gamma}$ with index
$\gamma=3.0$ (the dashed line in the lower panel of Figure
\ref{fig:profile}).

But IH95 also noted that a King profile was not necessarily a good fit
to their data.  The shape of the IH95 surface density
profile is similar to the model King profile out to distances of
$\sim30\arcmin$, beyond which their signal to noise becomes small.  In
order to overcome the signal to noise limitations, they averaged the
counts in several radial bins past the $30\arcmin$ limit (we reproduced
these points using the data in their Table 3 and plot them in our
Figure \ref{fig:profile}).  The average density in the $\sim36\arcmin$,
$\sim46\arcmin$, and $\sim56\arcmin$ bins of IH95 is
significantly higher than predicted by the King model.  The stars that
contribute to the deviation from the King profile fit are referred to
as ``extra-tidal stars'' by IH95, and they suggest these stars
may be indicative of ongoing tidal disruption of Ursa Minor by the
Galaxy.

Numerical simulations of tidally disturbed stellar systems in the outer
halo of the Galaxy \citep{kvj99b} predict that the surface density
profile should exhibit a ``break'' in the density fall-off, beyond
which the unbound stars begin to contribute more signal to the profile
than do the bound stars, eventually becoming the dominant population.
Beyond this break, the model predicts that the stellar density should
follow a shallow power law dropoff of $\Sigma(r) \sim r^{-1}$.  The
IH95 star count profile of Ursa Minor is quite similar to those derived
from the numerical simulations of \citet{kvj99b}.  However, the
profiles constructed using the Ursa Minor giant candidates identified
in this study do {\em not} show a sharp break in the profile followed
by a shallow dropoff in the stellar densities.  Instead, the profile
deviates slowly from the King profile (suggesting that an increased
tidal and/or core radius King profile compared to that of IH95 would
provide a better fit), and then it follows a $\Sigma(r) \sim r^{-3}$
power law decay in the outer regions that is steeper than the
predictions for the density fall-off of extratidal stars made by \citet
{kvj99b}.  The bright subsample shows the smoothest profile, however,
the two fainter subsamples perhaps show a break at roughly
$40\arcmin$.  The deviation in density near this potential break point
is not large compared to the error bars, though, so we hesitate to
conclude with any certainty that this is indeed a break as predicted in
the simulations of disrupting satellites.  The stellar densities in the
two fainter subsamples for radii past this potential break still follow
a $\sim r^{-3}$ power law, however.

Due to the east/west asymmetry and other morphological peculiarities
seen in Ursa Minor, it is possible that the usual methodology employed
to study the global profiles of dwarf galaxies is less appropriate in
the case of UMi.  The surface density profile derived from the global
spatial distribution of giants will not represent local features well.
To investigate possible variations induced by nonsymmetry, we have
derived radial surface density profiles for all UMi giant stars east of
the minor axis and also for all UMi giant stars west of the minor axis
(the adopted center of UMi is found between the two density peaks, thus
the primary density peak is found west of the minor axis and the
secondary peak is east of the minor axis).  The method employed was
identical to that used to create Figure \ref{fig:profile}, however, in
this case candidate UMi giant stars were counted in semi-ellipses on
either side of the minor axis.  Also, due to differing sample
completeness on either side of Ursa Minor, only those annuli that
completely fit within the survey boundaries are used in this analysis.

The morphological variations, evident in the surface density maps
(Figures \ref{fig:isodens} and \ref{fig:isodsmth}), are clearly
manifest in the comparison of the radial profiles constructed from
stars east and west of the minor axis.  There are two major differences
between the eastern and western radial profiles apparent in all three
panels of Figure \ref{fig:eastwest}.

\begin{itemize}

\item  The innermost point of the western profile is consistently
higher than the corresponding point on the eastern profile.  This
reflects the presence of the strong peak in the star counts in the western
half of UMi.  Within the core region (semi-major axis $\leq 13.6\arcmin$),
the density falls off more quickly on the western half than it does on 
the eastern half.  This is another way of saying the core region is more
centrally concentrated toward the western side.

\item  Outside of the core region (semi-major axis $\geq 20.4\arcmin$),
the three western profiles follow the roughly $r^{-3}$ power law decay
seen in the global profile (Figure \ref{fig:profile}).  However, in all
three eastern profiles, the $34\arcmin$ point is enhanced compared to
the corresponding point on the western profile, while the $40.8\arcmin$
point is suppressed.  The change in these two points makes the eastern
profiles appear to break, as seen in numerical simulations of tidally
disrupting systems \citep{kvj99b} and in observations of Carina
\citep{haloII}.  A more recent surface density profile for Ursa Minor
has been produced by \citet{md01}, and they also find a similar jump
between their $30\arcmin$ and $40\arcmin$ points, suggesting that the
feature seen in Figure \ref{fig:eastwest} is real (and also that our
sample of UMi stars, selected completely independently, traces the
same structures as does their sample).  For the two brighter samples
(top two panels of Figure \ref{fig:eastwest}), the power law index of
the points past the break on the eastern profiles is shallower ($\sim
r^{-2}$) than for these same points on the western profiles ($\sim
r^{-3}$).

\end{itemize}

Unfortunately, these differences between the eastern and western radial
profiles are at low statistical significance.  However, the eastern
profile suggests the possibility that, at least locally, the radial
profile of Ursa Minor breaks at $r = 34\arcmin$.  That this break
radius also corresponds to the break radius observed by IH95 suggests
the adoption of about $34\arcmin$ as the beginning of significant
contribution by unbound stars.  We may use this radius to estimate the
mass loss rate of Ursa Minor by the formalism of \citet{kvj99b}.  The
estimated mass loss rate quoted for Ursa Minor in \citet{kvj99b} of
32\% uses the value of $r_{break}$ from IH95.  If we adopt $r_{break} =
34.0\arcmin$ and $r_{xt} = 95.2\arcmin$ (the outermost annulus measured
with high statistical significance in the bright subsample), then from
the data in Table \ref{surfdens} and in Table 4 of \citet{kvj99b} we
find a fractional mass-loss rate of $df/dt_1 = 0.33$ Gyr$^{-1}$.  If
this stellar mass loss rate has been roughly constant over the lifetime
of the galaxy, then the mass of Ursa Minor was $0.67^{-N}$ times larger
N Gyr in the past.  If the central velocity dispersion for Ursa Minor
and its inferred $M/L$ ratio of 79 are correct, then the mass of this
dSph is currently $\sim 2 \times 10^7 M_{\sun}$ \citep{mm98}, and if
total mass loss tracks stellar mass loss, a constant $df/dt$ suggests
that UMi may have been $> 10^9$ solar masses 10 Gyr ago.  However, the
exact mass of Ursa Minor, the amount of mass loss by the dSph to the
halo, and the form (i.e., stars and/or dark matter) of the mass lost
depend strongly on input assumptions, and thus the exact mass of stars
and dark matter deposited in the Galactic halo by Ursa Minor is highly
uncertain.

Stars that are being removed from Ursa Minor will wind up in the
Galactic halo, and they may be detectable as an accreted population;
for example, by their kinematics \citep[cf. the moving group
of][]{srmetal96}.  The Galactic halo is made up predominantly of old,
metal-poor stars with little or no net rotation around the Galaxy.
However, there is evidence that the distant outer halo stars may have a
net retrograde rotation \citep{srm92a}, which can not occur in a
population formed during a monolithic collapse.  This retrograde
rotation has been confirmed \citep{srm92b, carney}, and, furthermore,
metallicity information suggests an additional difference between the
retrograde component of the halo:  The metallicity of the ``high halo''
retrograde stars peaks around $[\rm{Fe}/\rm{H}] \sim -2.0$, while the
mean metallicity for the ``low halo'' is $[\rm{Fe}/\rm{H}] \sim -1.6$.
Both \citet{srm92a} and \citet{carney} suggest that the retrograde
rotation may be attributable to accretion by the Galaxy of other
stellar systems, and the metallicity differences between the retrograde
stars and the majority of the halo perhaps provide support for this
hypothesis.  It has been argued \citep{una96} that the halo can only
contain a small percentage of stars accreted from the dSphs,
particularly if the typical accreted dSph had a stellar population
similar to the present day populations of Carina and Fornax, which have
dominant \textit{intermediate} age populations.  However, Ursa Minor
apparently contains only a single old stellar population
\citep[e.g.,][]{mm98, sf99} with a mean metallicity near
$[\rm{Fe}/\rm{H}] \sim -2.0$ \citep{shetrone01}.  Thus, those Ursa
Minor stars accreted into the halo will be virtually indistinguishable
from the Galactic ``high halo'' population.

Since Ursa Minor has a prominent population of BHB stars (unlike many
of the other dSphs), the disruption of UMi will contribute BHB stars to
the Galactic halo.  A large scale photographic study of the Galaxy
\citep{beers85} has been used to identify candidate blue ``field
horizontal branch'' (FHB) stars \citep[e.g.,][]{psb91, wil99}.  The
density of these stars on the sky is low, however, and the magnitude limit
of the photographic plates ($B \sim 16$) is such that these FHB stars
are mostly nearby and the outer halo is not sampled.  A catalogue of
more distant FHB stars has been created \citep{flynn}, however, it
covers far less area than the earlier study of \citet{beers85}, and it
only includes a few stars more distant than 20 kpc.  Thus, the overall
distribution of FHB stars in the {\em outer} halo remains unknown.
While the density of FHB stars in outer halo streams possibly can be
used to constrain the disruption history of UMi, the lack of a
well-defined, complete sample of distant FHB stars prevents a
comparison at this time.  

We note that the Sloan Digital Sky Survey \citep[SDSS;][]{york00} is
expected to provide a catalogue of outer halo FHB stars useful for
mapping the structure of the Galactic halo.  Some SDSS data (nearly 400
deg$^2$) on A-colored stars \citep{yanny00}, both FHB stars and blue
stragglers, have been used to map out substructure in the halo.
\citet{yanny00} determine the masses of two large substructures that
they have identified to be a few $\times 10^6 M_{\sun}$, and suggest
that they may be streamers produced by tidal disruption of dwarf
galaxies.  Using additional SDSS data on F-type stars,
\citet{newberg02} confirm the existence of the \citet{yanny00}
substructures, and associate them with streams from the
Sagittarius dwarf galaxy.  A similar search for FHB star streams
potentially associated with UMi is feasible when more SDSS data become
available.

\subsection{Implications for the Dark Matter Content of UMi}

The morphology and radial profile
of Ursa Minor appear to support the hypothesis that
this stellar system is being influenced strongly by the tidal field of
the Milky Way.  This accumulated evidence leads us to question the
widely accepted notion that the dSph is dark matter dominated.
\citet{burkdm97} has proposed that UMi can only be dark matter
dominated if the ``extratidal'' stars are not in fact extratidal; that
is, if the dark matter halo is very massive and much more spatially
extended than the stellar component of the dSph, and the stars
identified in this study and others as ``extratidal'' are bound within
the potential of the dark matter halo.  However, if the
\citet{burkdm97} model for the dSph is the correct one, it suggests
that the Milky Way tidal field should only be altering the morphology
of the dark matter component of UMi, since the stars must be
comfortably within the tidal radius of the dark matter halo.  In this
scenario, one expects that the stars will be distributed smoothly
within the dSph gravitational potential, the shape of which is
determined by the dominant dark matter component.  However, it is clear
that the stellar component of UMi is far from smooth, and this is a
problem for the models where the dSph is embedded in a massive,
extended dark matter halo model.

While the morphological evidence for ongoing tidal disruption of Ursa
Minor appears strong, there does not yet appear to be a satisfactory
explanation for producing an inflated velocity dispersion in a
disrupting system.  Although the dSph model of \citet{kk98} can produce
inflated velocity dispersions for unbound systems that extend along the
line of sight, this is not applicable to our sight line with Ursa
Minor, and it also predicts that the width of the horizontal branch
should be inflated due to the distance modulus variations of the
unbound stars found along the line of sight.  We can rule out any width
of the HB larger than $\sim0.3$ magnitudes (although the photometric
errors at the level of the HB are relatively large), which is smaller
than the $\sim1$ magnitude width seen in simulated data by
\citet{kk98}.  Also, while we have attributed the asymmetric morphology
of the Ursa Minor RGB and BHB stars to the tidal influence of the Milky
Way, it may be possible to construct a non-standard dark matter halo
model in which the morphology of the stellar population of Ursa Minor
could arise.  The current state of our knowledge of the extended
distribution of the stars associated with Ursa Minor alone does not
appear to allow us to rule out the presence of a large dark matter
component for this dSph.

Relying on the current data on UMi, we can however address the
accuracy of the reported values of $M/L$ for the dSph.  The standard
method for measuring $M/L$ requires an estimate of the total mass
derived from the core velocity dispersion, and an estimate of the total
luminosity estimated from observations.  Since the dSphs for the most
part cover large areas of the sky, it is difficult to 
determine their luminosities accurately.  For example, IH95 estimated the
total luminosity of Ursa Minor by using the King profile fit to define
an aperture expected to enclose 90\% of the light.  In their \S 7.1.1,
they point out that for UMi, the largest fraction of the error budget
in their $M/L$ estimate of $95 \pm 43$ is due to the error in
$L_{tot}$, which they calculate to be $2.0\pm0.9 \times 10^{5}
L_{\sun}$.  Clearly, this estimate hinges on the accuracy of the King
profile in defining the extent of Ursa Minor.  However, this profile
was fit to data that were background limited after $\sim 30\arcmin$.
Since the current sample of UMi stars has eliminated much of the
background, we are able to trace the dSph to radii of $\gtrsim
100\arcmin$ with reasonable signal-to-noise.  

Assuming that a King profile is a reasonable model for fitting the
profile of UMi (our data suggest this may not be true), we have used a
Maximum Likelihood technique (similar to the one presented in K98) to
fit the data points presented in Figure \ref{fig:profile}.  The best
fit King model to our $M_0 \leq 20.0$ sample of UMi RGB candidates has
$r_t = 77.9\arcmin \pm 8.9\arcmin$ and $r_c = 17.9\arcmin \pm
2.1\arcmin$ for an elliptical galaxy with $\theta =
49^{\circ}\pm1.6^{\circ}$ and $\epsilon = 0.54\pm0.02$.  The large
error bar on the tidal radius is due to the difficulty in fitting a
King model to these data.  However, our data are most consistent with a
large tidal radius due to the continuing steep decline in the surface
density seen out to $\sim100\arcmin$.  This increase in the total
extent of Ursa Minor also influences the total luminosity of the
system:  The total luminosity of the system is directly proportional to
the integral of the surface density profile.  Based on the King profile
fit to Ursa Minor presented here, we estimate that IH95 may have
underestimated the total luminosity of Ursa Minor by nearly a factor of
$\sim2.7$.  The parameters of the King fit (specifically the change in
concentration) also affect the mass estimate, but not as significantly
as they do the luminosity.  Based only on the change in the King
profile parameters, we calculate that the standard $M/L$ value for Ursa
Minor should be a factor of 2 smaller than previous estimates, i.e., 47
rather than 95.

It has been argued that no single effect can account for the large
inferred $M/L$ value for Ursa Minor.  For example, an anisotropic
velocity dispersion may affect the calculation of the virial mass.  It
may be unlikely that the velocity dispersion of UMi is significantly
anisotropic because this would require that the dSph be elongated along
the line of sight, which seems improbable based on its elongation in
the plane of the sky and on its narrow horizontal branch.  However, if
anisotropy is present to some degree in the system's velocity
dispersion, it may only inflate $M/L$ by a factor of $\sim3$
\citep{rt86}; not enough to produce $M/L \sim 100$ from a system with a
true $M/L \sim 3$.  Although this effect is often discounted since it
does not reduce the $M/L$ of UMi to a value expected for a system with
no dark matter, it is worth reconsidering now since anisotropy could
reduce $M/L \sim 47$ to $M/L \sim 16$.  That is, while we agree that it
is unlikely that one single physical effect has inflated the inferred
$M/L$ of Ursa Minor to 100, if several effects, such as underestimating
$L_{tot}$ and the repercussions from an anisotropy of the velocity
dispersion each contribute a factor of $2 - 3$ to the measured value of
$M/L$, it may be that Ursa Minor has a true $M/L$ that is much less
extreme when each of the ``minor'' effects on $M/L$ is correctly taken
into account.  We note that the Leo II dSph, which is located in the
distant outer halo and is presumably less affected by the tidal
influence of the Galaxy, has $M/L = 10$ \citep{mm98}.

The discussion presented here concerns the global $M/L$ ratio for Ursa
Minor.  An important assumption in the measurement of a global $M/L$
ratio is that mass follows light, which introduces significant
uncertainty into the process because it is not clear that mass follows
light in dSph galaxies.  For example, \citet{kleyna02} present models
for the shape of the dark matter halo in the Draco dSph, and they rule
out a mass follows light model for that dSph.  Because of the
uncertainty in the global distribution of dark matter in dSphs, the
$M/L$ ratio in the core is likely to be more robust than the global
value.  The central $M/L$ is insensitive to small changes in the total
luminosity, such as the one presented here, however the central $M/L$
is affected by the presence of substructure within the core.  Thus, our
observations do suggest that both the core and global $M/L$ ratios in
UMi should be reconsidered.

\section{A New Model for Ursa Minor}

The morphology and surface density profile of Ursa Minor derived in
previous studies have been used to argue that the stellar system is in
the process of losing stars to the Milky Way.  Based on accumulated
photometric observations, the ``standard'' model for Ursa Minor is that
it is a bound, possibly relaxed elliptical system of $50\arcmin$ major
axis tidal radius with ``extratidal'' stars found at semi-major axis
radii $\gtrsim 50\arcmin$ or so.  The time to destroy unbound
substructure in UMi by phase mixing is roughly 1 Gyr, and thus the
presence of substructure within the tidal radius is explained by
assuming that the clumps of stars are recent phenomena related to the
disruption of the system.

\begin{figure}
\plotone{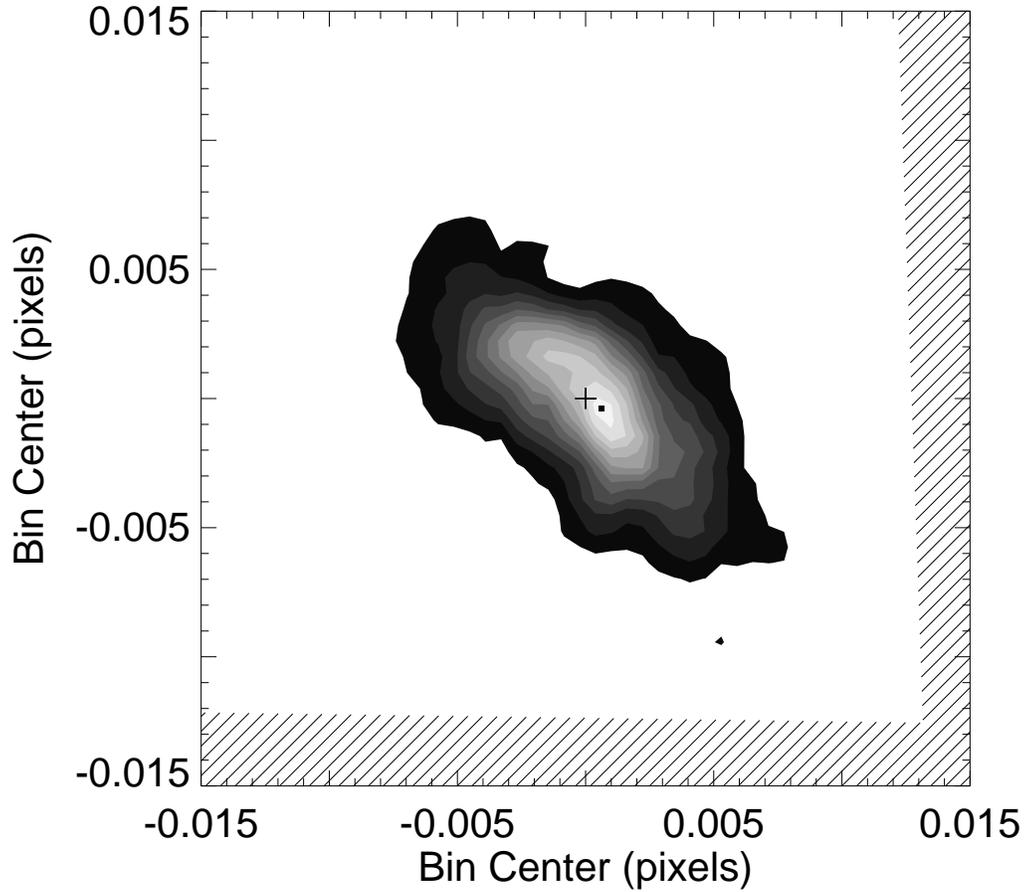}
\caption{Smoothed isodensity contours of Ursa Minor.  The single points
at each stellar position that were gridded to create Figure
\ref{fig:isodens} were replaced by two dimensional Gaussians centered
at the stellar position.  The ``smoothed'' stars were rebinned using a
finer, $100 \times 100$ pixel grid.  The counts in each pixel are
represented as a contour plot here.  The K98 center is
represented by the cross.  The center of symmetry of the ``ring'' of
UMi stars seen in the HST images of \citet{bd99} is plotted as a
filled square.  Note that this square is slightly offset to the
northeast of the densest region of Ursa Minor in the image presented
here.}
\label{fig:isodsmth}
\end{figure}

\begin{figure}
\plotone{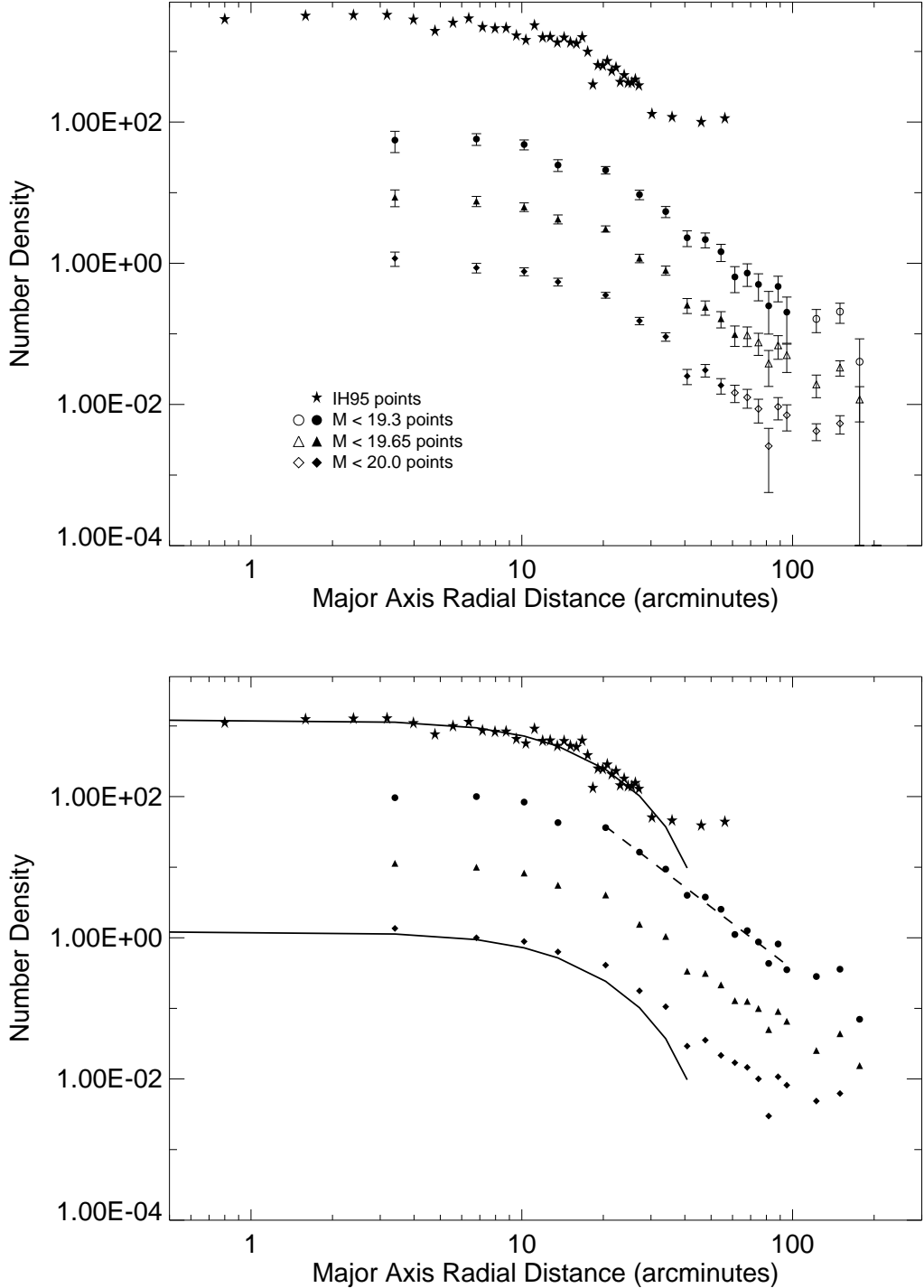}
\caption{Radial surface density profile of Ursa Minor.  In both panels,
the points have been offset vertically by a factor of 10 for ease of comparison.  The
upper panel represents the background subtracted stellar densities in
arcmin$^{-2}$ for the three magnitude-limited samples of giant
candidates presented in Figure \ref{fig:umingi}:  The filled circles
represent $M_0 \leq 19.3$ giants, the filled triangles represent $M_0
\leq 19.65$ giants, and the filled diamonds represent the $M_0 \leq
20.0$ giants.  The open symbols for each sample correspond to those
annuli that extend beyond the boundaries of our survey region and are
therefore not completely sampled. The filled stars are the background
subtracted densities from the Ursa Minor star counts of IH95.
The lower panel presents the normalized densities for comparison; the
densities have been normalized to 1.0 at $R \sim 6.8\arcmin$.  The
solid line is the King profile fit to the data by IH95, with
$r_t = 50.6\arcmin$ and $r_c = 15.8\arcmin$.  The dashed line is a
power law with index $\gamma = 3$ fit to the filled circles at radii
$\geq 20.4\arcmin$.}
\label{fig:profile}
\end{figure}

\begin{figure}
\plotone{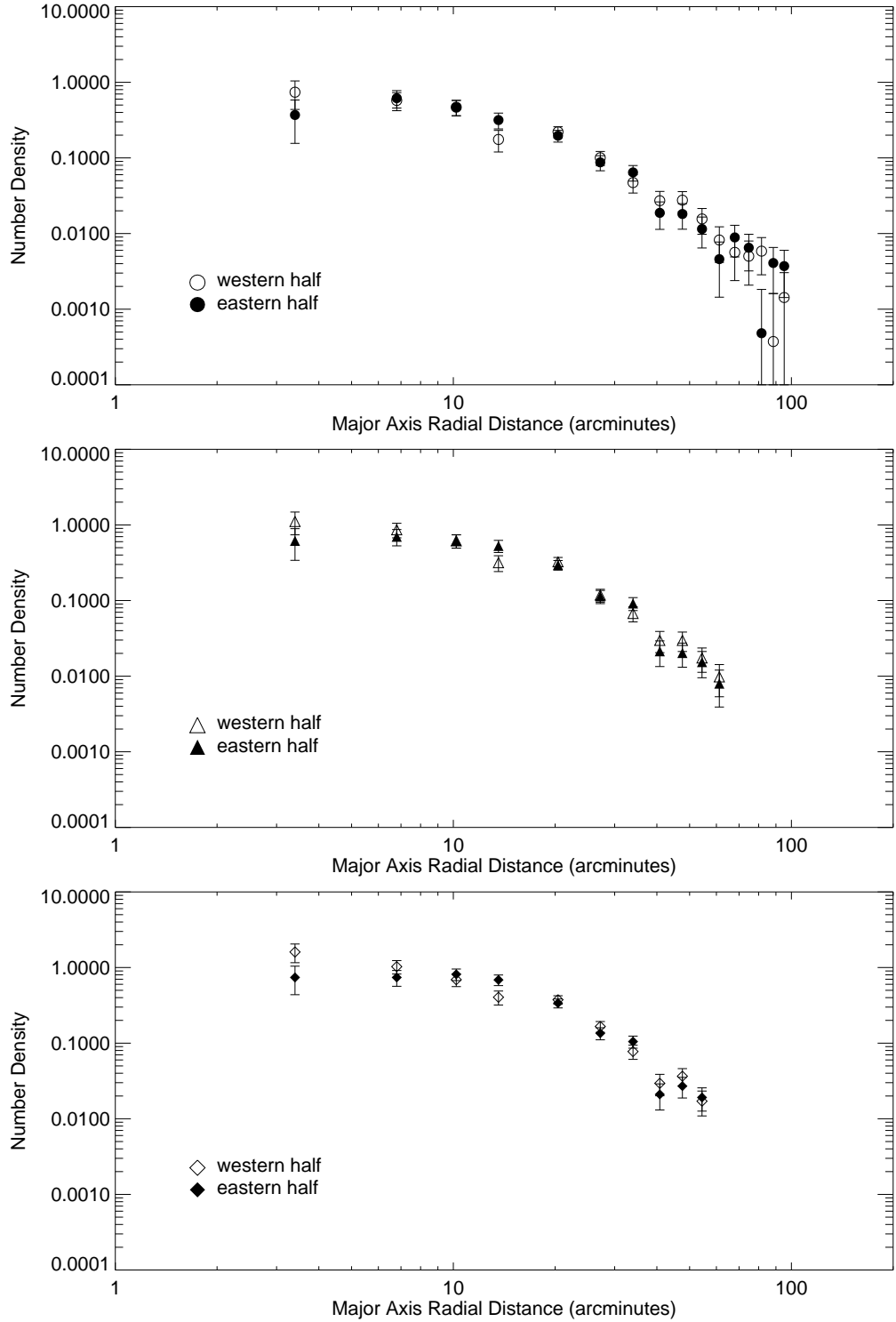}
\caption{Radial surface density profiles for eastern and western halves
of Ursa Minor.  In each panel, the open symbols represent the number
density of stars found in semi-ellipses west of the minor axis, while
the filled symbols represent the number density of stars found in
semi-ellipses east of the minor axis.  The upper panel (circles) is the
profile of the $M_0 \leq 19.3$ sample of UMi giant star candidates,
the middle panel (triangles) the $M_0 \leq 19.65$ sample, and the lower
panel (diamonds) is the $M_0 \leq 20.0$ sample.  Although the eastern
and western profiles are generally similar, two effects are seen:  Due to
the presence of the density peak in the western half of the galaxy,
the central density is enhanced and the subsequent decline is steeper
in the core region of the western profile than it is in the eastern
profile.  Also, the western profile decays with a $\Sigma(r) \sim
r^{-3}$ power law outside the core region, while the eastern profile
shows a break in the profile at $r=34.0\arcmin$, followed by a
shallower $\Sigma(r) \sim r^{-2}$ power law decay past the break
point.}
\label{fig:eastwest}
\end{figure}

The observations presented here lead us to propose a revision of the
standard model of Ursa Minor.  We postulate that the primary peak in
the surface density is the true core of Ursa Minor, analogous to the
suggestion that the ``globular cluster'' M54 is the nucleus or core of
the Sagittarius dwarf \citep{sl95}.  Assuming the peak isodensity
contours in Figures \ref{fig:isodens} and \ref{fig:isodsmth} define the
true core of Ursa Minor, then this suggests that the shape of the outer
isodensity contours reflects the tidal shaping of the bound stars
around this core and that the true ``center'' of the galaxy need not be
the center of symmetry of these outer contours.   We note that
independent observations that include fainter stars
of Ursa Minor not included in our sample \citep{md01b} confirm that the
central regions of the dSph contain a dense, globular cluster-like
region.

The simplest reason for identifying the offcenter primary density peak
seen in the isodensity plot of Ursa Minor as the core of the galaxy is
that it is the densest region in the galaxy; all theoretical models of
dSph structure (e.g., King models, power-law$+$core models, CDM halo
models, etc.) are constructed so that the highest density is found in
the center of the galaxy.  In addition to this simple argument,
however, there are several other reasons for identifying this structure
as the core of the galaxy:  (1)  Theoretical n-body simulations of dSph
galaxies evolving in a tidal field predict that the initially
spheroidal satellite becomes more elliptical as distance from the core
increases \citep{kvj02}.  Thus, at late times the simulated dSph still
has small ellipticity in the core, but increasing to $>0.5$ in the
outer regions.  The majority of the contours of Ursa Minor in Figure
\ref{fig:isodsmth} have ellipticity $>0.5$, however, the innermost
contours appear more spherical, and are consistent with
an ellipticity $< 0.5$.  (2) The
crossing time in Ursa Minor is short, $\sim10^7 - 10^8$ years, and
therefore substructure within the galaxy is expected to be erased on
timescales short compared to the Hubble time.  If the core of Ursa
Minor is indeed $>15\arcmin$ in radius (centered between the two
density peaks) and is well mixed, then the presence of an offcenter
density enhancement is difficult to explain, unless it is a recently
formed feature.  However, if we identify the highest density peak as
the core, then only those stars found within this much smaller region
are expected to be well mixed (and the smooth appearance of the contours
surrounding this peak suggests this is true) and it is reasonable to
expect to see substructure outside of the core, such as the secondary
density peak and the twisting of the outer contours.  (3)  In WF/PC2 images
of the core of UMi (centered on the position of highest stellar density
from the authors' earlier ground-based imaging) \citet{bd99} find a
``ring'' of stars; the ring has a central void, surrounded by a
circularly symmetric, dense distribution of stars.  However, a surface
density profile constructed using circular annuli centered at the ring
center finds the highest density at a radius of $\sim30\arcsec$.  This
offset is more consistent with the location of the center determined
from Figure \ref{fig:isodsmth}, and, indeed, in the image of chip WF3 in
\citet{bd99} there is an excess of stars at the edge of the chip near
where we predict the center of UMi to be.  Moreover, \citet{bd99}
suggest that the central region of UMi does not follow a King law, but
the best fit to their data has a core radius of $\lesssim 2.5\arcmin$,
rather than the $\sim 15\arcmin$ value derived previously.  

Identifying the peak in the UMi isodensity contours as the core of the
galaxy implies that the core radius is much smaller than previously
estimated and also that a larger fraction of the galaxy is susceptible
to influence by the Galactic potential than is often supposed.
Observations of another stellar system in the Milky Way halo support
these conclusions.  Recent discovery of extensive tidal tails
associated with the globular cluster Pal~5 \citep{oden01} provides a
system for comparison with Ursa Minor.  Although Pal~5 is a lower mass
system than UMi, the behavior of unbound, tidally stripped stars
depends on the shape of the Galactic potential and the satellite's
orbit rather than its initial structure \citep{kvj02}, so the
distribution of the unbound stars around Pal~5 is perhaps reasonable to
compare to the extended distribution of UMi stars if they, too, are
unbound.  Comparing the isodensity contours of Pal~5 \citep[Figure 2
of][]{oden01} to those of UMi, there appears to be a similar
``\textsf{S}-shaped'' morphology in both systems.  The bend in the
contours of Pal~5 is in the direction of the orbital path of the
system, as is the bend in the contours of Ursa Minor.  Although this
``\textsf{S}-shaped'' morphology appears to be related to the tidal
disruption of Pal~5 and, by analogy, therefore, to Ursa Minor, it is
unclear what is causing this bend.  N-body simulations of disrupting
systems show that the distribution of particles appears
\textsf{S}-shaped because the escaping particles initially leave
perpendicular to the orbit (i.e., along the direction of the tidal
force), but then bend in the direction of the orbital path as the
particles (stars) sort by orbital energy \citep[e.g.,][]{kvj02}.
However, this shape is {\em in the plane of the orbit}, that is, in
order to see it one needs to view the orbit face on.  Our line of sight
to Ursa Minor is nearly radial, and thus it is unlikely that the
\textsf{S} morphology seen in N-body simulations should be visible from
our point of view.

In order to test this assumption, we took N-body data from
\citet{kvj02} and projected the plane containing the particles of the
disrupted satellite along the orbital plane of Ursa Minor as determined
by the proper motion of \citet{kmc01u}.  As expected, the particles
that cause the usual, energy-sorting \textsf{S} morphology in the
orbital plane appear linear and symmetric when seen in projection, even
after accounting for the Sun's $\sim8$ kpc offset from the Galactic
center.  Although this suggests that the \textsf{S}-shape derived in
N-body simulations is not the source of the twisting seen in the
contours of UMi and Pal~5, since the angle between the line of sight to
UMi and its orbital plane ($\sim25^{\circ}$) is so similar to the angle
between the LOS to Pal~5 and its orbital plane ($\sim33^{\circ}$), the
physics responsible for creating this morphology in Pal~5 may be
responsible for the similar morphology seen in UMi.  Rotation or
tumbling of these two systems may be a viable method for the
\textsf{S}-shaped morphology to become visible from our line of sight,
however this remains to be investigated.  

We note that recent, detailed simulations (M. Odenkirchen 2002, private
communication) of the disruption of Pal~5 claim to reproduce the
\textsf{S}-shaped morphology of this globular cluster and confirm that
it is related to the transition by stars from radial escape to tangential drift
parallel to the cluster's orbit.  The results of these simulations show
that the bending is largest when viewing the orbital plane face-on, but
the projection onto the plane of the observer also shows it clearly, in
accordance with the observations.  

\section{Summary and Conclusions}

Using three color photometry, we have surveyed $\gtrsim9$ square
degrees in a region centered on the Ursa Minor dwarf spheroidal
galaxy.  The filters used are designed to allow efficient dwarf/giant
luminosity classification of all stars observed.  From among the
$>14,000$ objects measured that have stellar profiles and small
photometric errors, we have selected a sample of 788 candidate Ursa
Minor giant stars and 505 candidate Ursa Minor blue horizontal branch
stars.  A comparison of our catalogue of giants with spectroscopically
verified Ursa Minor members and non-members shows that so far we are
100\% accurate in separating contaminants and slightly less accurate
at recovering all member giants.

Among the candidate UMi stars, many lie beyond the nominal tidal
radius of Ursa Minor, as determined by previous studies of the dSph
(that assume the King limiting radius is the tidal radius).  These
``extratidal'' RGB stars (which may or may not be bound to the dSph)
lie in all directions and are seen to the limits of the surveyed region
(up to $\sim2.5 - 3.0$ degrees from the center).  Due to the intrinsic
faintness of the BHB stars, our detection of these stars is incomplete
outside of the central survey fields of Ursa Minor.  However, even
among this incomplete sample of BHB stars we find many extratidal
stars, and the spatial distribution of the BHB stars appears almost
identical to that of the RGB stars.

Analysis of the spatial distribution of the UMi giant candidates and
BHB candidates suggests that Ursa Minor has a peculiar morphology in
comparison to the majority of dwarf spheroidals.  Unlike most dSphs
where the highest density is found at the center of symmetry of the
outermost, high signal-to-noise isodensity contours, in UMi, two
off-center regions are found to be the areas with the highest stellar
density.  The western density peak is centrally concentrated and it is
less elliptical ($\epsilon \sim 0.25$) than the majority of the galaxy
($\epsilon=0.55$).  The second density peak has been seen by other
authors and is not an artifact of the data, but the excess of stars at
this location in UMi is of lower significance.  The stars that create
this peak are not concentrated, but are instead elongated in the
direction of the orbital path of UMi.  A smoothed representation of the
isodensity contours shows that the galaxy has a crescent-shaped, or
\textsf{S}-shaped, morphology.

In order to account for the variation in limiting magnitude across the
survey region, we divided the candidate UMi RGB stars into three
magnitude limited subsamples, $M_0 \leq 19.3$, $M_0 \leq 19.65$, and
$M_0 \leq 20.0$.  We derived surface density profiles of Ursa Minor by
counting RGB stars in concentric ellipses using the structural
parameters of K98.  In each case, the resulting profiles look
remarkably similar to each other at all radii and to the profile of UMi
derived by IH95 for radii $r <30\arcmin$ or so.  The surface density
profile is not fit well by previous King profile fits or by our own
King profile fit; the densities begin to deviate from the King profile
calculated by IH95 perhaps as close as $6.8\arcmin$ to the center, with
larger deviations seen at all points $r >20.4\arcmin$.  There may be a
``break'' seen in the surface density profile, but it is not as sharp
as the one seen in Carina, for example.   The points along the surface
density profile outside of $20.4\arcmin$ are fit best by a power law
with index $\sim -3$.  Due to the east/west asymmetry seen in the
isodensity contours, we derived surface density profiles separately for
the eastern and western halves of the dSph.  In contrast to the overall
profile and that of the western half, the profile of the eastern half
of the galaxy does appear to break at $34.0\arcmin$ (similar to IH95),
and the densities past this break follow a more shallow, $r^{-2}$ power
law past this point.  The radial profiles derived by IH95 and K98 have
high signal-to-noise only for radii $r <30\arcmin$, and thus the
discrepancy at large radii between the profile presented here and those
derived in these previous studies can be attributed to our ability to
study UMi at much lower stellar surface densities because we are mostly
free from contamination by foreground stars.

Based on the two-dimensional distribution of candidate Ursa Minor
stars, we conclude that this system is very likely undergoing
significant mass loss due to its tidal interaction with the Milky Way.
The morphology of the interior region of the galaxy that is usually
considered the core is inconsistent with expectations for a relaxed
system.  Instead, we propose that the smooth, fairly round density peak
west of the ``center'' of Ursa Minor is the core of the galaxy, and all
stars outside this roughly $2\arcmin$ radius region are asymmetrically
distributed around this core due to the tidal influence of the Milky
Way.  Our finding that the $M/L$ ratio of UMi may be as low as 16 is
more consistent with substantial mass loss than previous, large $M/L$
values.  However, we cannot rule out a model in which all of the RGB
and BHB stars identified in this study as associated stars are instead
bound within an extended, massive dark matter halo.  Further
investigation of the velocities of the extratidal stars is necessary to
determine if they are bound or unbound to the dSph.  The photometric
survey of Ursa Minor presented here was specifically undertaken in
order to identify prime candidates for spectroscopy; a spectroscopic
campaign to determine membership and to measure radial velocities for a
number of candidate Ursa Minor RGB stars is already underway.

\acknowledgements

CP wishes to acknowledge the generous support of the University of
Virginia Department of Astronomy during the completion of this work,
which is a significant portion of his Ph.D. thesis.  We note that the
referee, Tad Pryor, made numerous suggestions that resulted in a
significantly improved manuscript.  The authors also wish to
acknowledge helpful discussions with Kathryn Johnston and Andrea
Schweitzer.  This work benefitted from the authors participation in the
discussions and debate at the 2001 Ringberg Workshop on The Lowest-Mass
Galaxies and Constraints for Dark Matter, organized by Eva Grebel.  We
wish to acknowledge support for this work from NSF CAREER award
AST-9702521, the David and Lucile Packard Foundation, and a Cottrell
Scholar Award from the Research Corporation.  CP also acknowledges
partial support from NSF award AST 00-71223.

\clearpage

\clearpage

\begin{deluxetable}{rccccccccc}
\tablecaption{UMi Candidate RGB and BHB Star Photometry\tablenotemark{a}}
\tablehead{\colhead{ID} & \colhead{$\alpha$\tablenotemark{b}} & 
\colhead{$\delta$\tablenotemark{b}} & \colhead{$M$} 
& \colhead{$\sigma_{M}$} & \colhead{$T_2$} & \colhead{$\sigma_{T_2}$} & 
\colhead{$DDO51$} & \colhead{$\sigma_{DDO51}$} & \colhead{Type}}
\startdata
     15 & 15:11:45.13 & 67:21:45.4 & 19.80 &  0.03 & 18.63 &  0.02 & 19.71 &  0.03 & rgb   \\
  32088 & 15:10:17.65 & 67:20:15.4 & 18.40 &  0.01 & 17.16 &  0.01 & 18.37 &  0.02 & rgb   \\
1800035 & 15:09:18.40 & 66:26:32.3 & 18.45 &  0.02 & 17.15 &  0.01 & 18.41 &  0.02 & rgb   \\      
      7 & 15:11:51.56 & 67:26:46.9 & 19.85 &  0.03 & 19.78 &  0.06 & 19.86 &  0.03 & bhb   \\
 202047 & 15:16:07.53 & 67:47:21.5 & 19.96 &  0.06 & 19.43 &  0.09 & 19.86 &  0.06 & bhb   \\
\enddata

\tablenotetext{a}{The complete version of this table is in the electronic edition of
the Journal.  The printed edition contains only a sample.}
\tablenotetext{b}{J2000.0}

\label{photom}
\end{deluxetable}

\begin{deluxetable}{lccc}
\tablewidth{250pt}
\tablecaption{Number of Color-Color Selected Giants in RGB Bounding Box}
\tablehead{ & \multicolumn{3}{c}{Number of Giants}
\\ \colhead{$\Delta M$} & \colhead{$M_0 < 19.3$} &
\colhead{$M_0 < 19.65$} & \colhead{$M_0 < 20.0$}}
\startdata
-0.33 &  267 & 392 &  486 \\
-0.66 &  121 & 223 &  311 \\
-0.99 &  \phn47 & \phn98 &  183 \\
\cline{2-2}
-1.32 &  \phn33 & \phn50 &  \phn83 \\
\cline{3-3}
-1.65 &  \phn34 & \phn41 &  \phn50 \\
\cline{4-4}
-1.98 &  \phn24 & \phn34 &  \phn36 \\
-2.31 &  \phn25 & \phn28 &  \phn29 \\
-2.64 &  \phn23 & \phn31 &  \phn22 \\
\cline{2-4}
-2.97 &  \phn25 & \phn28 &  \phn23 \\
\enddata

\label{bkgrndtab}
\end{deluxetable}

\begin{deluxetable}{lccccc}
\tablewidth{365pt}
\tablecaption{Ursa Minor Giant Candidate Counts and Background Levels}
\tablehead{\colhead{Magnitude} & \colhead{UMi Giant} & \colhead{Area} &
\colhead{Extratidal\tablenotemark{a}} &
\colhead{Area\tablenotemark{b}} & \colhead{Background} \\
\colhead{Limit} & \colhead{Counts} &
\colhead{deg$^2$} &
\colhead{Counts} & \colhead{deg$^2$} & \colhead{deg$^{-2}$} }
\startdata
$M_0 \leq 19.3$ & 393 & 9.06 & 100 & 8.06 & $3.1\pm0.6$ \\
$M_0 \leq 19.65$ & 540 & 8.31 & 139 &  7.31 & $4.0\pm0.7$ \\
$M_0 \leq 20.0$ & 600 & 5.56 & 119 & 4.56 & $5.2\pm1.0$ \\
\enddata

\tablenotetext{a}{Stars found outside the IH95 tidal radius}
\tablenotetext{b}{An ellipse drawn from either the IH95 or
K98 structural parameters has area of $\sim1$ deg$^2$.}

\label{subsamps}
\end{deluxetable}

\begin{deluxetable}{cccccccc}
\tablewidth{420pt}
\tablecaption{Star Counts of Ursa Minor Giant Candidates in Elliptical Annuli}
\tablehead{ & & \multicolumn{2}{c}{$M_0 \leq 19.3$} &
\multicolumn{2}{c}{$M_0 \leq 19.65$} & \multicolumn{2}{c}{$M_0 \leq 20.0$} \\
\colhead{$r_{in}$} & \colhead{$r_{out}$} &
\colhead{Area} & \colhead{Counts} & \colhead{Area} & \colhead{Counts} &
\colhead{Area} & \colhead{Counts} \\
\colhead{arcmin} & \colhead{arcmin} & \colhead{arcmin$^2$} &  &
\colhead{arcmin$^2$} & & \colhead{arcmin$^2$} & }
\startdata
\phn\phn0.0 & \phn\phn3.4 & \phn\phn16.2 & \phn9 & \phn\phn16.2 & \phn14 & 
\phn\phn16.2 & \phn19 \\
\phn\phn3.4 & \phn\phn6.8 & \phn\phn48.6 & 28 & \phn\phn48.6 & \phn37 & 
\phn\phn48.6 & \phn42 \\
\phn\phn6.8 & \phn10.2 & \phn\phn81.0 & 39 & \phn\phn81.0 & \phn51 & \phn\phn81.0
 & \phn62 \\
\phn10.2 & \phn13.6 & \phn113.4 & 28 & \phn113.4 & \phn48 & \phn113.4 & \phn62 \\
\phn13.6 & \phn20.4 & \phn323.9 & 68 & \phn323.9 & 100 & \phn323.9 & 115 \\
\phn20.4 & \phn27.2 & \phn453.5 & 43 & \phn453.5 & \phn54 & \phn453.5 & \phn70 \\
\phn27.2 & \phn34.0 & \phn583.1 & 32 & \phn583.1 & \phn47 & \phn583.1 & \phn54 \\
\phn34.0 & \phn40.8 & \phn712.7 & 17 & \phn712.7 & \phn19 & \phn712.7 & \phn19 \\
\phn40.8 & \phn47.6 & \phn842.3 & 19 & \phn842.3 & \phn21 & \phn842.3 & \phn27 \\
\phn47.6 & \phn54.4 & \phn971.8 & 15 & \phn971.8 & \phn17 & \phn971.8 & \phn19 \\
\phn54.4 & \phn61.2 & 1101.4 & \phn8 & 1101.4 & \phn12 & \phn993.6 & \phn16 \\
\phn61.2 & \phn68.0 & 1231.0 & 10 & 1217.3 & \phn13 & \phn996.9 & \phn14 \\
\phn68.0 & \phn74.8 & 1360.6 & \phn8 & 1265.6 & \phn11 & \phn989.3 & \phn10 \\
\phn74.8 & \phn81.6 & 1490.2 & \phn5 & 1219.1 & \phn\phn6 & \phn994.9 & 
\phn\phn4 \\
\phn81.6 & \phn88.4 & 1619.7 & \phn9 & 1249.5 & \phn10 & 1026.2 & \phn11 \\
\phn88.4 & \phn95.2 & 1730.0 & \phn5 & 1309.5 & \phn\phn8 & 1059.5 & \phn\phn9 \\
\phn95.2 & 122.4 & 7232.9 & 18 & 6583.8 & \phn20 & 4427.9 & \phn25 \\
122.4 & 149.6 & 6837.0 & 20 & 6747.8 & \phn30 & 2788.3 & \phn19 \\
149.6 & 176.8 & 6328.6 & \phn8 & 6118.5 & \phn14 & 2875.5 & \phn\phn2 \\
176.8 & 204.0 & 7704.5 & \phn1 & 7320.7 & \phn\phn4 & 3860.9 &  \phn\phn1 \\
\enddata
\label{surfdens}
\end{deluxetable}

\end{document}